\documentclass[sigconf, nonacm, pdfa]{acmart}

\newcommand\vldbdoi{10.14778/3742728.3742752}
\newcommand\vldbpages{2616-2625}
\newcommand\vldbvolume{18}
\newcommand\vldbissue{8}
\newcommand\vldbyear{2025}
\newcommand\vldbauthors{\authors}
\newcommand\vldbtitle{\shorttitle} 
\newcommand\vldbavailabilityurl{https://github.com/PKU-DAIR/Hetu}
\newcommand\vldbpagestyle{empty} 

\newif\ifappendix
\appendixtrue

\usepackage{amsmath}
\usepackage{amsthm}
\usepackage{amssymb}
\usepackage{mathtools}
\usepackage{bm}
\usepackage{bbm}
\usepackage{booktabs}
\usepackage{xspace}
\usepackage{color,soul}
\usepackage{multirow}
\usepackage{multicol}
\usepackage{caption}
\usepackage{enumitem}

\usepackage{microtype}
\usepackage{graphicx}
\usepackage{subfigure}
\usepackage{booktabs}
\usepackage{hyperref}
\usepackage{nicefrac}
\usepackage[normalem]{ulem}
\usepackage{tikz}
\usetikzlibrary{trees}
\usepackage{float}
\usepackage{subfigure}
\usepackage[ruled,noend,linesnumbered,vlined]{algorithm2e}
\usepackage{xcolor,colortbl}
\usepackage{hhline}
\usepackage{tablefootnote}

\usepackage{listings}
\usepackage{pifont}

\definecolor{myred}{rgb}{1.0,0.7,0.8}
\definecolor{mygreen}{RGB}{0,166,0}
\definecolor{lightgreen}{RGB}{180, 255, 180}
\definecolor{myorange}{RGB}{252,107,4}
\definecolor{darkgreen}{RGB}{0,153,102}
\definecolor{lightblue}{rgb}{0.53, 0.81, 0.92}
\definecolor{lightgray}{gray}{0.9}

\newcommand{\mysubsubsection}[1]{{\vspace{0.25em}\noindent\ul{\textit{\textbf{#1.\xspace}}}}}

\newcommand{\myvspace}[1]{\vspace{#1}}

\DeclareMathOperator*{\argmin}{arg\,min}

\newcommand{\system}{{\texttt{LobRA}}\xspace}

\newtheorem{theorem}{Theorem}

\newtheorem{definition}{Definition}
\newtheorem{observation}{Observation}
\newtheorem{property}{Property}
\newtheorem{assumption}{Assumption}

\newtheorem*{statement*}{\statementnumber}
\providecommand{\statementnumber}{}


\begin{document}

\title{\system: Multi-tenant Fine-tuning over Heterogeneous Data}

\author{Sheng Lin}
\authornote{Equal contribution.}
\authornote{School of Computer Science \& Key Lab of High Confidence Software Technologies (MOE), Peking University}
\affiliation{
\institution{Peking University}
\country{}
}
\email{linsh@stu.pku.edu.cn}

\author{Fangcheng Fu}
\authornotemark[1]
\authornote{School of Artificial Intelligence, Shanghai Jiao Tong University}
\affiliation{
\institution{Shanghai Jiao Tong University}
\country{}
}
\email{ccchengff@gmail.com}

\author{Haoyang Li}
\authornotemark[2]
\affiliation{
\institution{Peking University}
\country{}
}
\email{lihaoyang@stu.pku.edu.cn}

\author{Hao Ge}
\authornotemark[2]
\affiliation{
\institution{Peking University}
\country{}
}
\email{gehao@stu.pku.edu.cn}

\author{Xuanyu Wang}
\authornotemark[2]
\affiliation{
\institution{Peking University}
\country{}
}
\email{wxyz0001@pku.edu.cn}

\author{Jiawen Niu}
\authornotemark[2]
\affiliation{
\institution{Peking University}
\country{}
}
\email{niujiawen705@stu.pku.edu.cn}

\author{Yaofeng Tu}
\affiliation{
\institution{ZTE Corporation}
\country{}
}
\email{tu.yaofeng@zte.com.cn}

\author{Bin Cui}
\authornotemark[2]
\authornote{Institute of Computational Social Science, Peking University (Qingdao)}
\affiliation{
\institution{Peking University}
\country{}
}
\email{bin.cui@pku.edu.cn}

\begin{abstract}
With the breakthrough of Transformer-based pre-trained models, the demand for fine-tuning (FT) to adapt the base pre-trained models to downstream applications continues to grow, so it is essential for service providers to reduce the cost of processing FT requests. 
Low-rank adaption (LoRA) is a widely used FT technique that only trains small-scale adapters and keeps the base model unaltered, conveying the possibility of processing multiple FT tasks by jointly training different LoRA adapters with a shared base model. 

Nevertheless, through in-depth analysis, we reveal the efficiency of joint FT is dampened by two heterogeneity issues in the training data --- the sequence length variation and skewness. 
To tackle these issues, we develop \system, a brand new framework that supports processing multiple FT tasks by jointly training LoRA adapters. 
Two innovative designs are introduced. 
Firstly, \system deploys the FT replicas (i.e., model replicas for FT) with heterogeneous resource usages and parallel configurations, matching the diverse workloads caused by the sequence length variation. 
Secondly, for each training step, \system takes account of the sequence length skewness and dispatches the training data among the heterogeneous FT replicas to achieve workload balance. 
We conduct experiments to assess the performance of \system, validating that it significantly reduces the GPU seconds required for joint FT by 45.03\%-60.67\%. 
\end{abstract}

\maketitle

\pagestyle{\vldbpagestyle}
\begingroup\small\noindent\raggedright\textbf{PVLDB Reference Format:}\\
\vldbauthors. \vldbtitle. PVLDB, \vldbvolume(\vldbissue): \vldbpages, \vldbyear.\\
\href{https://doi.org/\vldbdoi}{doi:\vldbdoi}
\endgroup
\begingroup
\renewcommand\thefootnote{}\footnote{\noindent
This work is licensed under the Creative Commons BY-NC-ND 4.0 International License. Visit \url{https://creativecommons.org/licenses/by-nc-nd/4.0/} to view a copy of this license. For any use beyond those covered by this license, obtain permission by emailing \href{mailto:info@vldb.org}{info@vldb.org}. Copyright is held by the owner/author(s). Publication rights licensed to the VLDB Endowment. \\
\raggedright Proceedings of the VLDB Endowment, Vol. \vldbvolume, No. \vldbissue\ %
ISSN 2150-8097. \\
\href{https://doi.org/\vldbdoi}{doi:\vldbdoi} \\
}\addtocounter{footnote}{-1}\endgroup

\ifdefempty{\vldbavailabilityurl}{}{
\vspace{.3cm}
\begingroup\small\noindent\raggedright\textbf{PVLDB Artifact Availability:}\\
The source code, data, and/or other artifacts have been made available at \url{\vldbavailabilityurl}.
\endgroup
}

\section{Introduction}
\label{sec:intro}

Transformer-based~\cite{attn} pre-trained models, represented by Large Language Models (LLMs)~\cite{chatgpt,gpt4,llama2,qwen2}, have fueled an ever-increasing demand for their deployment in various applications such as chatbot assistants~\cite{dialogpt,roller2020recipes}, machine translation~\cite{back_translation,multilingual_translate}, summarization~\cite{pegasus_abs_summarize,text_summarize_encoder}, database tuning~\cite{gptuner_db_tuning_gpt,dbot_db_diagnosis_llm,dbgpt}, and more~\cite{dl4code_gen_survey,multi_behavior_rec,deng2023deep,li2022end}.
Since pre-trained models are mostly trained with general data, it is a common practice to leverage a fine-tuning (FT) process~\cite{ft1,ft2,ft3}, which further trains the pre-trained model with domain-specific data to adapt the model to the target applications.

Driven by the advancement of LLMs, many technology companies are enhancing their business strategies through the adoption of the Model as a Service (MaaS) paradigm~\cite{maas}. 
To this end, providing FT services becomes essential. 
For example, a technology company may own a closed-source pre-trained model and allow users to upload their private or domain-specific datasets for FT~\cite{openai_ft_docs}. 
For another example, cloud service providers would also offer FT services using popular open-source pre-trained models~\cite{databricks_ft_docs,snowflake_ft_docs}. 
As a result, given the diverse downstream applications, there would be many FT requests on the basis of the same pre-trained model, and consequently, it is of great value to reduce the cost associated with the FT requests over the same base model.

Meanwhile, FT services exhibit specific characteristics that are divergent from other kinds of services like model inference. 
For one thing, each user can submit multiple FT requests simultaneously, in order to build various domain-specific models with the same base model or to evaluate different dataset mixtures to see which gives the best performance~\cite{dual_stage_mixed_ft,tulu}. 
For another, compared to the inference scenario, FT requests arrive far less frequently (e.g.,~\cite{ymir_project_dataset_and_workload} reported an average of around 8.5 FT tasks per hour and some of them arrive simultaneously), and it takes significantly longer to process an FT request (tens of minutes to hours in practice), so both the request arrival rate and departure rate are much lower. 
Consequently, there would be a batch of FT requests that co-exist for long durations, and the batch does not change frequently. 
Given these characteristics, this work focuses on \textit{how to reduce the cost of jointly executing a batch of FT tasks over the same base model}.

\begin{figure}[!t]
\centering
\includegraphics[width=.8\linewidth]{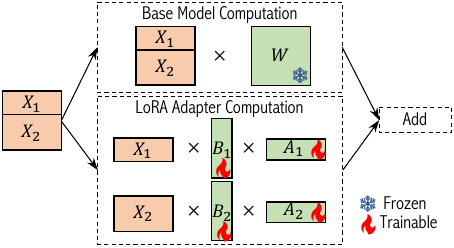}
\myvspace{-8pt}
\caption{\small{An illustration of the fusion of different LoRA adapters.}}
\label{fig:lora_batching}
\myvspace{-15pt}
\end{figure}

Owing to the astonishing model sizes and the scarcity of hardware accelerators (typically, GPUs), low-rank adaptation (LoRA)~\cite{lora,lora_llm_survey} has become one of the most widely used and effective FT techniques. 
In essence, LoRA only trains small-scale adapters that consist of much fewer parameters than the base model, significantly reducing the FT cost. 
Since LoRA keeps the base model unaltered, it conveys the opportunity to share the same base model across multiple FT tasks rather than maintaining individual model replica(s) for each one. 
As shown in Figure~\ref{fig:lora_batching}, in each training step, we can fuse the input data from different tasks so that the computation of the base model can be fused into a batched operation whilst the computation of multiple LoRA adapters can be supported by customized operations. 
In fact, co-serving multiple LoRA adapters has been widely investigated for LLM inference~\cite{pets,s_lora,punica,lora_inlaid}. 
Inspired by this, this work focuses on how to efficiently carry out multiple FT tasks as a joint FT task by fusing multiple LoRA adapters.

Nevertheless, we experienced unsatisfactory efficiency when we na\"ively leveraged the batch fusion idea to support joint FT. 
After an in-depth investigation, we found that this is because of two heterogeneity issues of the training data for joint FT. 

\begin{figure}[!t]
\centering
\includegraphics[width=\linewidth]{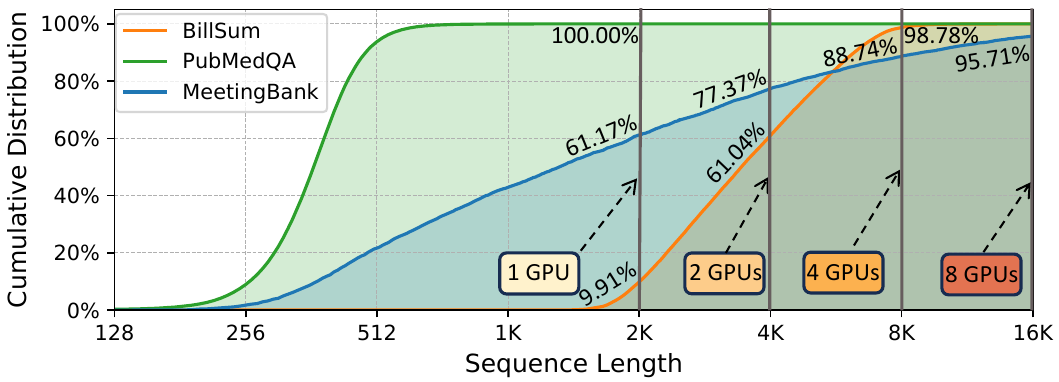}
\myvspace{-20pt}
\caption{\small{Cumulative distributions of sequence lengths of three FT datasets. The ``$n$ GPU(s)'' indicates that we need $n$ A100-40GB GPU(s) to process data with the corresponding sequence length without out-of-memory errors when fine-tuning the Llama2-7B model.}}
\label{fig:seqlen_cdf}
\myvspace{-20pt}
\end{figure}

The first is \ul{\textit{the sequence length variation among tasks}}. Specifically, since Transformer models take sequences as input, the training data (sequences) inevitably vary in length. As shown in Figure~\ref{fig:seqlen_cdf}, the sequence length distributions are substantially divergent across FT tasks. 
This is reasonable as the training data of a few tasks (e.g., summarization) are usually much longer than the others (e.g., question answering). 
Meanwhile, since the memory consumption of FT is linear w.r.t. the lengths~\cite{flash_attn,flash_attn_v2,memo_long_context}, it requires different numbers of GPUs to support the processing of sequences with different levels of lengths. 
Nevertheless, existing works overlook the discrepancy in resource demands, and straightforwardly deploy the FT replicas (i.e., model replicas for FT) in the same way for all training data, which conforms to the number of GPUs required for the highest sequence length. 
This would lead to efficiency degradation as the communication cost becomes higher for most training data. 

To cope with this problem, this work proposes the idea of \ul{\textit{heterogeneous FT replicas}}, which deploys the FT replicas with varying resource usages and parallel configurations. 
Thus, we accelerate the processing of short sequences with FT replicas with low model parallel degrees, whilst avoiding out-of-memory errors by dispatching the long sequences to FT replicas with high model parallel degrees. 

However, \ul{\textit{the skewness in sequence lengths}}, the second heterogeneity issue of FT data, raises another hurdle. 
As shown in Figure~\ref{fig:seqlen_cdf} again, most of the training data (sequences) are relatively short, which is a natural characteristic of human texts~\cite{context_length_analysis}. 
If we simply dispatch the training data to FT replicas according to their lengths, the workloads would be extremely imbalanced among the replicas. 
That is to say, FT replicas with low model parallel degrees must process much more training data, leading to heavier workloads compared to those with high parallel degrees. 
Since FT replicas must synchronize the parameters of LoRA adapters for every training step, it inevitably results in idle periods for some FT replicas.

We address this issue by developing a \ul{\textit{workload-balanced data dispatching}} technique. 
This is inspired by the fact that short sequences can also be processed by FT replicas with high model parallel degrees. 
Thus, we adjust the data dispatching so that the workloads of FT replicas with low model parallel degrees can be migrated to those with high parallel degrees, achieving workload balance.

Putting them together, this work presents \system, a multi-tenant FT framework by manufacturing multiple LoRA adapters concurrently. 
To tackle the data heterogeneity issues, \system innovatively introduces the deployment of heterogeneous FT replicas and workload-balanced data dispatching to accelerate the joint FT process. 
In summary, this work makes the following contributions.
\begin{itemize}[leftmargin=*]
\item 
We first anatomize different design choices of joint FT over heterogeneous training data, revealing the necessity of supporting heterogeneous model deployment and workload-balanced data dispatching. 
Based on the anatomy, we formulate a joint optimization problem to co-optimize these two factors.

\item 
Subsequently, we propose a two-stage decomposition of the joint optimization problem for practical joint FT. 
On instantiation, the first-stage problem determines the heterogeneous model deployment plan that is optimal in expectation. 
Then, for each training step of the joint FT process, we derive the workload-balanced data dispatching plan by solving the second-stage problem. 

\item 
We evaluate the performance of \system by fine-tuning LLMs with up to 70B parameters and more than 10 tasks over 64 GPUs. Empirical results demonstrate that \system effectively reduces the GPU seconds required for joint FT by 45.03\%-60.67\%.
\end{itemize}

\section{Background and Related Works}
\label{sec:pre}


\subsection{Fine-tuning over Variable-length Data}
\label{sec:pre_ft_varlen}

\mysubsubsection{Fine-tuning based on LoRA}
Fine-tuning (FT) is an essential process to adapt the pre-trained model to the target domain~\cite{ft1,ft2,ft3}. 
Low-rank adaption (LoRA) and its variants~\cite{lora,lora_llm_survey} only fine-tune small-scale adapters rather than the base model itself, reducing resource demands substantially. 
For a model weight matrix $W \in \mathbb{R}^{in \times out}$, LoRA trains two low-rank matrices $A \in \mathbb{R}^{r \times out}, B \in \mathbb{R}^{in \times r}$ ($r \ll in, out$) and computes $XW + XBA$.
By doing so, it approximates the change to the weight matrix as $\Delta W \approx BA$.

As LoRA does not alter the base model, previous works have proposed the co-serving of multiple LoRA adapters for inference~\cite{pets,s_lora,punica,lora_inlaid}, and \citet{flexllm} further considered the scenario of inference and FT simultaneously. 
However, processing multiple FT tasks is not their primary focus. 
\citet{mlora_aspen} and \citet{sched_pricing_multi_lora} considered fusing multiple LoRA adapters for joint FT. 
Nevertheless, they carry out the joint FT task by na\"ively batching the input data, 
overlooking the data heterogeneity mentioned in \S\ref{sec:intro}. 
There are also many works that compose LoRA adapters for better performance~\cite{mixture_of_lora_experts,lora_compose_image_gen,multi_direct_lora,lorahub,multi_lora_multi_task}, which is orthogonal to our work.

\begin{figure}[!t]
\centering
\includegraphics[width=0.9\linewidth]{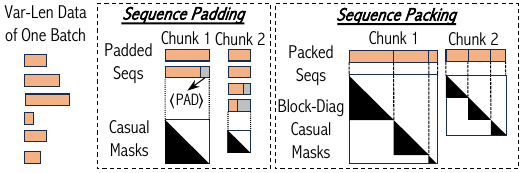}
\myvspace{-8pt}
\caption{\small{Illustration of applying sequence padding and packing to variable-length data of one batch. Sequence padding uses the special token $\langle$PAD$\rangle$ to ensure sequences within the same chunk are the same length. Sequence packing concatenates sequences together and uses the block-diagonal casual masks to avoid cross-contamination.}}
\label{fig:padding_and_packing}
\myvspace{-10pt}
\end{figure}

\mysubsubsection{Processing Variable-length Data}
In each training step, given a batch (a.k.a. mini-batch) of data, they are with diverse lengths due to the variable-length nature of sequences. 
Meanwhile, as GPU memory is limited, it is usually infeasible to process the batch at once. 
Thus, it is common to re-organize the batch into smaller chunks (a.k.a. micro-batches), process each sequentially, and accumulate model gradients computed from all chunks for model update. 

Figure~\ref{fig:padding_and_packing} shows two commonly used data re-organization techniques. 
Sequence padding sorts the data according to their lengths and takes those with similar lengths when crafting each chunk. 
Within each chunk, it adds special tokens (i.e. $\langle$PAD$\rangle$) to ensure a unique length. 
Sequence packing concatenates sequences for each chunk, eliminating the need for padding tokens. 
Meanwhile, it adjusts the casual mask to be block-diagonal 
to avoid cross-contamination among the sequences that are packed together~\cite{seq_packing}.

In theory, packing provisions better training efficiency. 
However, \citet{long_align} conducted experiments on FT tasks and observed that padding and packing exhibit comparable training efficiency, whilst training with padding usually achieves better model quality, which is because packing introduces biases to the contributions of different data. 
Thus, padding and packing have their pros and cons, and the choice between them is an open question.
In this work, we assume padding is employed. 
Nevertheless, it is noteworthy that the proposed designs can also be applied when packing is employed.

Given the variable-length phenomenon, there are also several works investigating more variants of packing and padding~\cite{icp,seq_pack_compare,fewer_trunc}. 
However, none of these works have considered processing data of different lengths with different resource usages and parallel configurations. 
As a result, they are orthogonal to our work.

\begin{figure*}[!t]
\centering
\includegraphics[width=\linewidth]{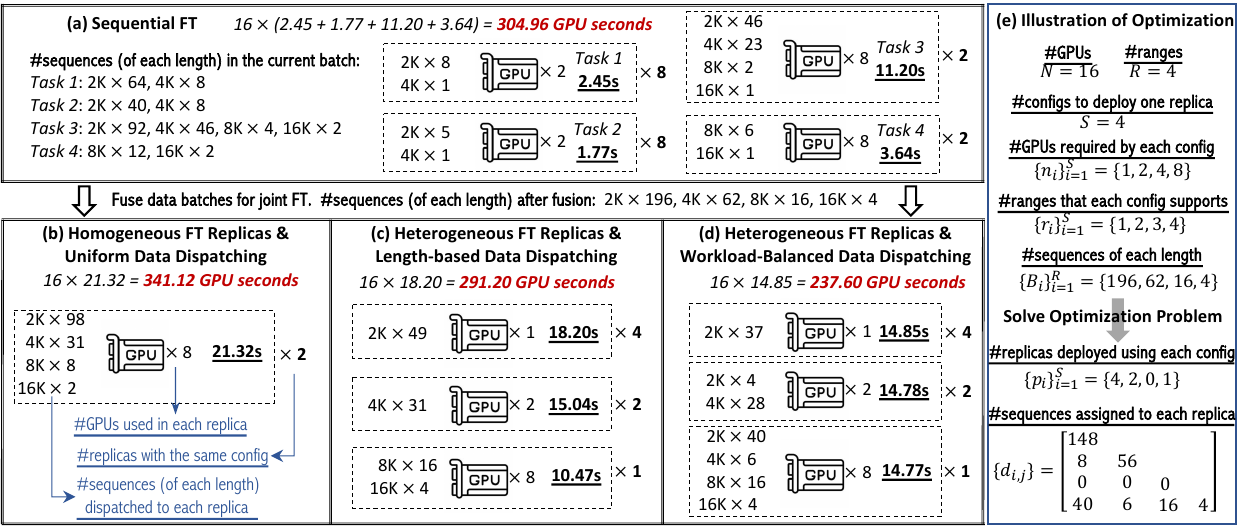}
\myvspace{-15pt}
\caption{\small{An example of 4 FT tasks with four different approaches, where (a) denotes fine-tuning the 4 tasks sequentially, whilst (b)-(d) present three different designs discussed in \S\ref{sec:design_anatomy}. We focus on the total GPU seconds required to run one training step for each task. (e) illustrates the inputs and decision variables of Equation~{\eqref{eq:origin_prob}} based on (d).}}
\label{fig:design_anatomy}
\myvspace{-10pt}
\end{figure*}

\subsection{Parallel Configurations in Model Training}
\label{sec:pre_parallel_config}

\mysubsubsection{Model Parallel}
There are two prevalent forms of model parallel, namely tensor parallel (TP)~\cite{megatron_1,megatron_3} and pipeline parallel (PP)~\cite{gpipe,pipedream,pipedream_flush,pipeline_survey}. 
Since TP and PP have different pros and cons, it is common to combine them for better efficiency, which is also known as hybrid model parallel~\cite{megatron_2,flexflow_soap,alpa,galvatron,galvatron_bmw,piper,MiCS}. 
In essence, model parallel distributes the model across GPUs to reduce the memory occupied by the model itself, sparing the space for intermediate results in training. 
Since the memory consumed by intermediate results is linear w.r.t. the summed lengths in each chunk~\cite{flash_attn,flash_attn_v2,memo_long_context}, if we wish to support longer sequences, we usually need to increase the model parallel degree, yet at the price of larger communication overhead.

\mysubsubsection{Model Replication (Data Parallel)}
Besides model parallel, a model can be replicated into multiple replicas. 
The input data are usually evenly dispatched to these replicas for concurrent processing (a.k.a. data parallel~\cite{pytorch_ddp,pytorch_fsdp,osdp,bagua,angelptm,cai2023adaptive}). 
To ensure the consistency of model parameters, it is necessary to synchronize the model gradients (or parameters) among the replicas for every training step, so balancing the workload across replicas is important.

\begin{table}[!t]
\centering
\caption{\small{Frequently used notation throughout this work.}}
\myvspace{-10pt}
\label{tab:notations}
\small
\begin{tabular}{|c|l|}
\hline
$N$ & The number of available GPUs. \\
\hline
$R$ & The number of sequence length ranges, i.e., the training data\\&of each batch are divided into $R$ buckets. \\
\hline
$S$ & The number of candidate parallel configurations. \\
\hline
$\mathcal{S}_i$ & The $i$-th candidate parallel configuration. \\
\hline
$n_i$ & The number of GPUs needed by $\mathcal{S}_i$ to deploy one FT replica. \\
\hline
$p_i$ & The number of FT replicas deployed with $\mathcal{S}_i$. \\
\hline
$r_i$ & The number of sequence length ranges that $\mathcal{S}_i$ supports,\\&i.e., FT replicas with $\mathcal{S}_i$ support processing sequences\\&in the first $r_i$ ranges without out-of-memory errors. \\
\hline
$d_{i,j}$ & The number of training data (sequences) in the $j$-th range \\&that are assigned to the FT replicas with $\mathcal{S}_i$. \\
\hline
\end{tabular}
\myvspace{-15pt}
\end{table}

\mysubsubsection{Cost Model of Fine-tuning Replicas}
In this work, we call each model replica a \textit{fine-tuning (FT) replica}. 
Each replica is associated with a parallel configuration that describes how it is parallelized. 

Given a training task, previous works generally develop cost modeling for running time and memory consumption, so that they can deduce how to parallelize the model~\cite{galvatron,galvatron_bmw,alpa,flexflow_soap,oobleck}. 
However, they only support the same parallel configuration for all model replicas and do not consider fine-tuning with variable-length data.

Fortunately, we can borrow ideas from previous works to build the cost model for variable-length data, which involves profiling the time cost of essential modules and estimating the running time according to the critical path of the training workflow. 
In addition, since the memory consumption is linear w.r.t. the summed lengths in each chunk~\cite{flash_attn,flash_attn_v2,memo_long_context}, we can easily profile the maximum supported sequence length for each kind of parallel configuration. 

Throughout this work, we divide the sequence length into $R$ non-overlapping ranges so that the variable-length data can be sorted into $R$ buckets. We assume there are $S$ candidate parallel configurations $\{\mathcal{S}_i\}_{i=1}^{S}$, where the $i$-th candidate requires $n_i$ GPUs to deploy one FT replica and supports processing sequences in the first $r_i$ ranges ($r_i \leq R$) without out-of-memory errors. 
Subsequently, we denote $T(\{d_{\cdot,j}\}_{j=1}^{r_\cdot}; \mathcal{S}_\cdot)$ as the time cost\footnote{Note that our work is applicable as long as the time cost function is linear w.r.t. $d_{\cdot,j}$.} of an FT replica associated with $\mathcal{S}_\cdot$, where $d_{\cdot,j}$ represents how many sequences in the $j$-th range are dispatched to this FT replica.
Due to the space constraint, we leave the details of our cost model in \ifappendix{Appendix~\ref{sec:appendix_time_cost_model}}\else{Appendix D}\fi ~\cite{lobra_appendix}.

\section{Design Anatomy}
\label{sec:design_anatomy}

This section anatomizes several design choices for joint FT, aiming to minimize the total GPU seconds needed to run one training step per task. 
To help readers better understand, we provide an example of four FT tasks in Figure~\ref{fig:design_anatomy}.
Figure~\ref{fig:design_anatomy}(a) depicts a sequential FT process that executes the four FT tasks one by one, whilst Figure~\ref{fig:design_anatomy}(b)-(d) illustrates three different design choices of joint FT. 

\mysubsubsection{Na\"ive Design: Homogeneous FT Replicas and Uniform Data Dispatching}
As shown in Figure~\ref{fig:design_anatomy}(b), an intuitive solution for joint FT is to instantiate the FT replicas with the same parallel configuration, fuse the data batches from all tasks together, and evenly dispatch them to these homogeneous FT replicas. 
By doing so, it is obvious that the training workloads are balanced across the replicas. 
However, the end-to-end performance of such an approach is unsatisfactory due to \ul{\textit{the sequence length variation among tasks}}.

Specifically, unlike well-structured tabular datasets, sequences inherently have diverse lengths. Moreover, the sequence lengths of different tasks are significantly divergent. 
Such a variation results in different memory consumption, thereby calling for different trade-offs between memory reduction and training efficiency. 
As a result, for tasks with long sequences, since their memory consumption is high, it is common to increase the model parallel degree (which requires more GPUs for one replica) to avoid out-of-memory errors, whilst incurring higher communication costs. 
In contrast, for tasks with short sequences, we can leverage a lower model parallel degree (which requires fewer GPUs for one replica) for better efficiency. 

Since the na\"ive approach uses the same parallel configuration for all replicas, the configuration must be able to accommodate the memory consumption of the \textit{longest} sequences to avoid out-of-memory errors. 
For instance, each FT replica in Figure~\ref{fig:design_anatomy}(b) occupies 8 GPUs, whilst only 2 GPUs are necessary if we consider the first two FT tasks individually in Figure~\ref{fig:design_anatomy}(a). 
Thus, their processing time would be significantly prolonged, making this approach ineffective.

\begin{figure*}[!t]
\centering
\includegraphics[width=.8\linewidth]{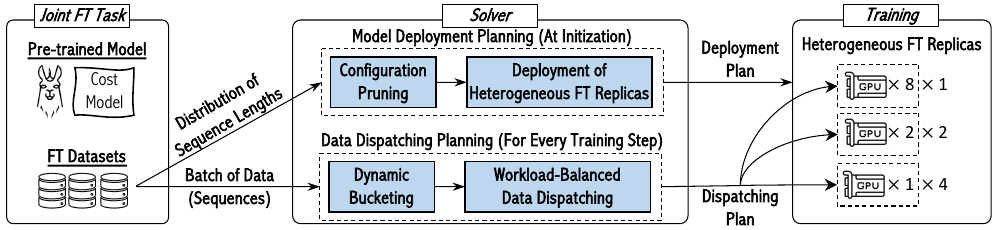}
\myvspace{-5pt}
\caption{\small{Overview of \system. To start the joint FT task, given the base model and the sequence length distribution of the FT datasets, \system determines the deployment plan of FT replicas that minimizes the running time in expectation. 
During the FT process, for each training step, \system analyzes how the corresponding batch of data should be dispatched among the FT replicas in order to achieve workload balance.}}
\label{fig:overview}
\myvspace{-10pt}
\end{figure*}

\mysubsubsection{Better Design: Heterogeneous FT Replicas and Length-based Data Dispatching}
Based on the discussion above, a better design is to leverage heterogeneous FT replicas for variable-length sequences. 
As exemplified in Figure~\ref{fig:design_anatomy}(c), we can instantiate the replicas with non-unique parallel configurations, classify the training data into buckets according to their lengths, and dispatch each bucket to the most suitable replica(s). 
By this means, from the perspective of each sequence, it can be processed by the most efficient configuration.
Nevertheless, such a solution suffers from the workload imbalance problem caused by \ul{\textit{the skewness in sequence lengths}}. 

To elaborate, in real-world corpora, most sequences are relatively short, and there are very few sequences that are significantly longer than others. 
Figure~\ref{fig:seqlen_cdf} presents the sequence length distributions of several FT datasets --- more than half of the sequences are shorter than 2K, whilst only a few are longer than 8K. 
In fact, this is reasonable in human texts and similar observations have also been reported on extremely large-scale pre-training corpora~\cite{context_length_analysis,llm_datasets_survey}. 

Back to the discussion about Figure~\ref{fig:design_anatomy}(c), given the skewness issue, the replicas with low model parallel degrees would receive a large portion of training data, whilst the other replicas would receive very few. 
As synchronization is needed for model update, replicas must idly wait for the slowest one(s). 
Worse still, a higher model parallel degree requires more GPUs for each replica, so there is a huge waste of computational resources. 
For the example in Figure~\ref{fig:design_anatomy}(c), the 8 GPUs in the third replica sit idle for approximately 42\% of the time during the joint FT process (10.47 vs. 18.20 seconds).

\mysubsubsection{Optimized Design: Heterogeneous FT Replicas and Workload-Balanced Data Dispatching}
Inspired by this, we propose to adopt the combination of heterogeneous FT replicas and workload-balanced data dispatching. 
The rationale is that replicas with high model parallel degrees can also process short sequences, so we can dispatch short sequences to more kinds of replicas. 
As depicted in Figure~\ref{fig:design_anatomy}(d), by doing so, we can strike a good balance among the heterogeneous FT replicas and improve end-to-end efficiency.

Given a batch of training data, let $B_j$ be the number of sequences fallen into the $j$-th bucket ($j \in [1, R]$). 
There are two key factors. 
Firstly, to deploy the heterogeneous FT replicas, we need to select the suitable parallel configurations and the number of replicas for each selected configuration.
Secondly, for each batch, we need to dispatch the training data to minimize the running time of the slowest replica. 
We formulate the following optimization problem:
\begin{equation}
\small
\begin{aligned}
\label{eq:origin_prob}
\argmin_{p_i, d_{i,j} \in \mathbb{N}_0 \text{ for } i \in [1, S], j \in [1, r_i]}
& \;\;\; \max_{i \in [1, S]} 
T\left( \left\{ \left\lceil {d_{i,j}}/{p_i} \right\rceil \right\}_{j=1}^{r_i}; \mathcal{S}_i \right) \\
\text{s.t.} 
& \sum_{i \in \{i \vert r_i \geq j\}} d_{i,j} = B_j \text{ for } \forall j \in [1, R] \\
& d_{i,j} \leq B_j \times p_i \text{ for } \forall i \in [1, S], j \in [1, r_i] \\
& \sum_{i=1}^{S} p_i \times n_i \leq N 
\end{aligned}
\end{equation}
where $p_i$ represents the number of FT replicas that are deployed with $\mathcal{S}_i$ (i.e., the $i$-th parallel configuration, and $p_i = 0$ indicates $\mathcal{S}_i$ is not selected for deployment), and $d_{i,j}$ represents how many sequences in the $j$-th bucket are assigned to the $p_i$ replica(s) with $\mathcal{S}_i$ for processing. 
$T(\cdot; \cdot)$ denotes the running time of the $p_i$ replica(s) with $\mathcal{S}_i$ given the dedicated sequences. 
The first constraint ensures all sequences are processed. 
The second constraint requires no sequences will be assigned if a configuration is not selected (i.e., $d_{i,j}=0$ as long as $p_i=0$). 
The third constraint ensures the FT replicas can be instantiated using the available GPUs.

We illustrate an example of the optimization problem with Figure~{\ref{fig:design_anatomy}}(d). 
There are $N=16$ GPUs and we divide the sequences into $R=4$ buckets based on their lengths. 
To deploy one replica, there are $S=4$ candidate configurations, requiring $\{n_i\}_{i=1}^S = \{1,2,4,8\}$ GPU(s) and supporting sequences in the first $\{r_i\}_{i=1}^S = \{1,2,3,4\}$ bucket(s), respectively. 
In the current fused batch, the numbers of sequences fallen into the buckets are $\{B_j\}_{j=1}^R = \{196, 62, 16, 4\}$. 
Then, by solving the optimization problem, the numbers of deployed replicas with the configurations are $\{p_i\}_{i=1}^S = \{4,2,0,1\}$, and the data dispatching is shown in Figure~{\ref{fig:design_anatomy}(e).}

\section{\system}
\label{sec:method}

\subsection{Overview}
\label{sec:method_overview}

Although Equation~\eqref{eq:origin_prob} co-optimizes the model deployment and data dispatching, solving it for every training step is impractical. 
For one thing, across different steps, the best model deployment plan may vary, which implies that we would need to reconfigure the model partitioning. 
Given the substantial model sizes, this is prohibitively time-consuming. 
For another, solving the problem takes longer than the training of one step (as evaluated in \S\ref{sec:expr_ablation}), making it infeasible to solve it for every step.
To tackle these obstacles, we propose a two-stage decomposition as depicted in Figure~\ref{fig:overview}. 
\begin{itemize}[leftmargin=*]
\item 
The first stage (\S\ref{sec:method_deployment}) produces the deployment plan of heterogeneous FT replicas. 
It is only done once at the initialization of the joint FT task, eliminating the need for reconfiguration of model deployment as well as the expensive solving cost for every step. 
\item 
Given the deployed FT replicas, the second stage (\S\ref{sec:method_dispatching}) only deduces the optimal data dispatching, which is fast and can be invoked for every step. 
This enables us to dynamically adapt the data dispatching to the randomly drawn batches during training. 
\end{itemize}

\subsection{Deployment of Heterogeneous FT Replicas}
\label{sec:method_deployment}

\mysubsubsection{Problem Formulation}
To instantiate the joint FT task, it is essential to determine how to deploy the heterogeneous FT replicas. 
However, according to Equation~\eqref{eq:origin_prob}, the optimal deployment plan is relevant to how the sequences are distributed across the buckets (i.e., $\{B_j\}_{j=1}^{R}$).
To address the discrepancy, we manage to utilize the overall distribution of sequence lengths of the FT datasets. 
In particular, we re-write Equation~\eqref{eq:origin_prob} as follows:
\begin{equation}
\small
\begin{aligned}
\label{eq:model_prob}
\argmin_{p_i, d_{i,j} \in \mathbb{N}_0 \text{ for } i \in [1, S], j \in [1, r_i]}
& \;\;\; \max_{i \in [1, S]} 
T\left( \left\{ \left\lceil {d_{i,j}}/{p_i} \right\rceil \right\}_{j=1}^{r_i}; \mathcal{S}_i \right) \\
\text{s.t.} 
& \sum_{i \in \{i \vert r_i \geq j\}} d_{i,j} \geq B \times f_{j} \text{ for } \forall j \in [1, R] \\
& d_{i,j} \leq B_j \times p_i \text{ for } \forall i \in [1, S], j \in [1, r_i] \\
& \sum_{i=1}^{S} p_i \times n_i \leq N 
\end{aligned}
\end{equation}
where $B$ denotes the batch size of the joint FT task, and $f_j$ denotes the percentage of sequences fallen into the $j$-th bucket. 
Solving Equation~\eqref{eq:model_prob} returns the model deployment plan ($p_i$) and data dispatching plan ($d_{i,j}$) that is optimal in expectation. 
Then, \system instantiates the FT replicas based on the model deployment plan (whilst the data dispatching plan will be omitted). 

Note that since $f_j$'s are not integers, we have inequality constraints on $d_{i,j}$ in Equation~{\eqref{eq:model_prob}}. 
Formally speaking, the solutions to Equation~{\eqref{eq:model_prob}} may have $\sum_{j} \sum_{i} d_{i,j} = B + R$, whilst the solutions to Equation~{\eqref{eq:origin_prob}} follow $\sum_{j} \sum_{i} d_{i,j} = B$, implying a difference. 
Nevertheless, after solving Equation~{\eqref{eq:model_prob}}, only $p_i$'s are used for model deployment, whilst $d_{i,j}$'s are omitted. 
Combining with the fact that $B \gg R$, the difference between Equation~{\eqref{eq:origin_prob}} and Equation~{\eqref{eq:model_prob}} is minor, and it does not affect the effectiveness of our work.

\mysubsubsection{Configuration Pruning}
Due to the term $d_{i,j}/p_i$, Equation~\eqref{eq:model_prob} is a mixed-integer non-linear programming (MINLP) problem, which is a notorious combinatorial optimization problem. 
Although libraries like SCIP~\cite{scip} support solving MINLP problems, it is very time-consuming, especially when there are many decision variables (i.e., $d_{i,j}$ and $p_i$). 
Specifically, the number of decision variables of Equation~{\eqref{eq:model_prob}} is $S + \sum_{i=1}^S r_i \leq S + S \times R$. 
Thus, the solving cost of Equation~\eqref{eq:model_prob} is highly related to $S$, as $R$ is not large in practice (e.g., we set $R=16$ in our experiments). 
To speed up the solving process, we introduce two heuristics to filter the candidate configurations.
Due to the space constraint, we briefly introduce the rationale below and refer interested readers to \ifappendix{Appendix~\ref{sec:appendix_config_pruning}}\else{Appendix A}\fi ~\cite{lobra_appendix} for more details.
\begin{itemize}[leftmargin=*]
\item 
Firstly, we only consider a small set of configurations (rather than covering all possible ones) for problem-solving. 
In essence, if two configurations consume the same number of GPUs (i.e., same $n$) and one of them is consistently less efficient than the other, then it will never be selected for deployment. 
Besides, since the model architecture of each pre-trained model is fixed, we conduct offline benchmarking and propose the candidates in advance, without affecting the online problem-solving. 

\item 
Secondly, although there is no closed-form solution for MINLP problems, it is possible to heuristically estimate the lower bound of training efficiency given a deployment plan (i.e., $\{p_i\}_{i=1}^{S}$). 
The reason is that, given the aforementioned benchmarking, we can estimate how the running time of two FT replicas would change after we migrate some training data between them. 
Thus, given a deployment plan, we treat the length-based dispatching as a starting point and estimate the lower bound of workload-balanced dispatching. 
Then we filter out deployment plans that are predicted to be inefficient, which shrinks the solution space.
\end{itemize}
Based on these two heuristics, Equation~{\eqref{eq:model_prob}} can be solved efficiently and accurately. 
For one thing, when the number of GPUs is relatively small (e.g., 16-32 GPUs), the heuristics do not affect the achieved solutions (i.e., the achieved solutions are the same as those achieved without pruning), yet accelerate the solving process substantially.
For another, when there are more GPUs, solving without pruning fails to finish within an hour, whilst solving with pruning only takes several minutes. 
Since Equation~{\eqref{eq:model_prob}} is only solved at initialization, the solving time is worthwhile given the speedup in the joint FT process. 
In \ifappendix{Appendix~\ref{sec:appendix_config_pruning} and Appendix~\ref{sec:appendix_more_expr_results}}\else{Appendix A and Appendix B.2}\fi ~\cite{lobra_appendix}, we have provided more details and empirical results.

\begin{figure}[!t]
\centering
\includegraphics[width=0.8\linewidth]{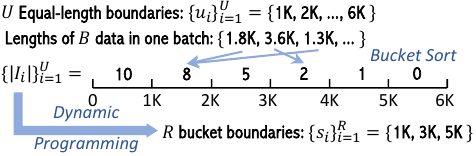}
\myvspace{-10pt}
\caption{\small{Illustration of dynamic bucketing.}}
\label{fig:dynamic_bucketing}
\myvspace{-15pt}
\end{figure}

\subsection{Workload-Balanced Data Dispatching}
\label{sec:method_dispatching}

\mysubsubsection{Problem Formulation}
After the heterogeneous FT replicas are deployed, each training step of the joint FT process involves randomly drawing a batch of training data, and feeding them to the FT replicas. 
Due to the randomness in batch sampling, the number of sequences that fall into each bucket may be divergent across different steps. 
Denote $p^*_i$ as the optimal solution achieved by Equation~\eqref{eq:model_prob}, which describes the model deployment plan of the heterogeneous FT replicas, we formulate the following optimization problem to minimize the time cost of each training step.
\begin{equation}
\small
\begin{aligned}
\label{eq:data_prob}
\argmin_{d_{i,j} \in \mathbb{N}_0 \text{ for } i \in [1, S], j \in [1, r_i]} 
& \;\;\; \max_{i \in [1, S]} 
T\left( \left\{ \left\lceil {d_{i,j}}/{p^*_i} \right\rceil \right\}_{j=1}^{r_i}; \mathcal{S}_i \right) \\
\text{s.t.} 
& \sum_{i \in \{i \vert r_i \geq j\}} d_{i,j} = B_j \text{ for } \forall j \in [1, R] \\
& d_{i,j} \leq B_j \times p^*_i \text{ for } \forall i \in [1, S], j \in [1, r_i] 
\end{aligned}
\end{equation}
Since $\{p^*_i\}_{i=1}^{S}$ are constants rather than decision variables, the optimization problem in Equation~\eqref{eq:data_prob} is an integer linear programming (ILP) problem, which can be efficiently solved via existing libraries like PuLP~\cite{pulp} and SCIP~\cite{scip}. 
The number of decision variables in Equation~{\eqref{eq:data_prob}} is $\sum_{i=1}^S r_i \leq S \times R$. 
However, since we can remove the configurations that are not selected for deployment (i.e., whose $p_i^*$ is zero), there are very few decision variables in practice. 
(For instance, there are only 3-5 selected configurations in our evaluation.)
Thus, as we will evaluate in \S\ref{sec:expr_ablation}, the solving process is fast and can be fully overlapped by the training of previous step(s).

\mysubsubsection{Dynamic Bucketing}
Till now, we assume that the boundaries of buckets, denoted as $\{s_i\}_{i=1}^{R}$, are pre-defined and fixed throughout the joint FT process. 
However, due to the randomness in batch sampling, the optimal boundaries vary across different training steps ---
as each sequence must be padded to the closest boundary, using fixed boundaries would lead to undesirable padding. 
To address this problem, we develop a dynamic bucketing approach to facilitate the adaption to the sequence length distribution during training.

As depicted in Figure~{\ref{fig:dynamic_bucketing}}, our approach starts from $U$ pre-defined boundaries $\{u_i\}_{i=1}^{U}$, which partition the range of sequence length into $U$ intervals (in practice, we consider equal-length division $\{256, 512, \cdots\}$). 
Given a batch of $B$ sequences, our approach bucket sorts them by lengths and determines $R$ boundaries ($R \leq U$) to form $R$ buckets that minimize the padding via dynamic programming.

Denote $\mathcal{I}_i$ as the set of indices of sequences fallen into the $i$-th interval, and $\textup{State}_{i,j}$ as the minimized padding when bucketing the first $i$ intervals into $j$ buckets. 
We have the following initial state and state transition expressions for dynamic programming:
\begin{equation}
\small
\begin{aligned}
& \textup{State}_{0,j} = 0, \ \forall j \in [0, R], 
\;\;\;\;\;\;
\textup{State}_{i,0} = +\infty, \ \forall i \in [1, U], 
\\
& \textup{State}_{i+1,j+1} = 
\min_{i^\prime \in [0,i]} 
\left\{
\textup{State}_{i^\prime,j} + 
\sum_{i^{\prime\prime} = i^\prime + 1}^{i}
\left( \left\lvert \mathcal{I}_{i^{\prime\prime}} \right\rvert \times (u_{i+1} - u_{i^{\prime\prime}}) \right)
\right\}
\end{aligned}
\end{equation}
By computing $\textup{State}_{U, R}$, we achieve the optimal boundaries for a given batch of data\footnote{The total number of padding tokens is $\textup{State}_{U, R} + \sum_{i=1}^{U} \sum_{k \in \mathcal{I}_{i}} (u_i - s_k)$, where the second term denotes the padding needed inside each interval, which is a constant.}.
The time complexity of the dynamic programming is merely $O(B + R U^2)$, introducing negligible overhead.\footnote{We default $R$ as 16 (sensitivity experiment provided in \ifappendix{Appendix~\ref{sec:appendix_more_expr_results}}\else{Appendix B.2}\fi ~\cite{lobra_appendix}) and ignore empty intervals in our implementation, so the term $RU^2$ is small in practice.}

Note that the bucketing approach is also applied when solving the deployment of FT replicas in Equation~\eqref{eq:model_prob}. 
In particular, at the initialization of the joint FT task, we randomly sample a large number ($100 \times B$ by default) of training data and perform the bucketing to determine the boundaries for solving Equation~\eqref{eq:model_prob}.

\begin{figure}[!t]
\centering
\includegraphics[width=\linewidth]{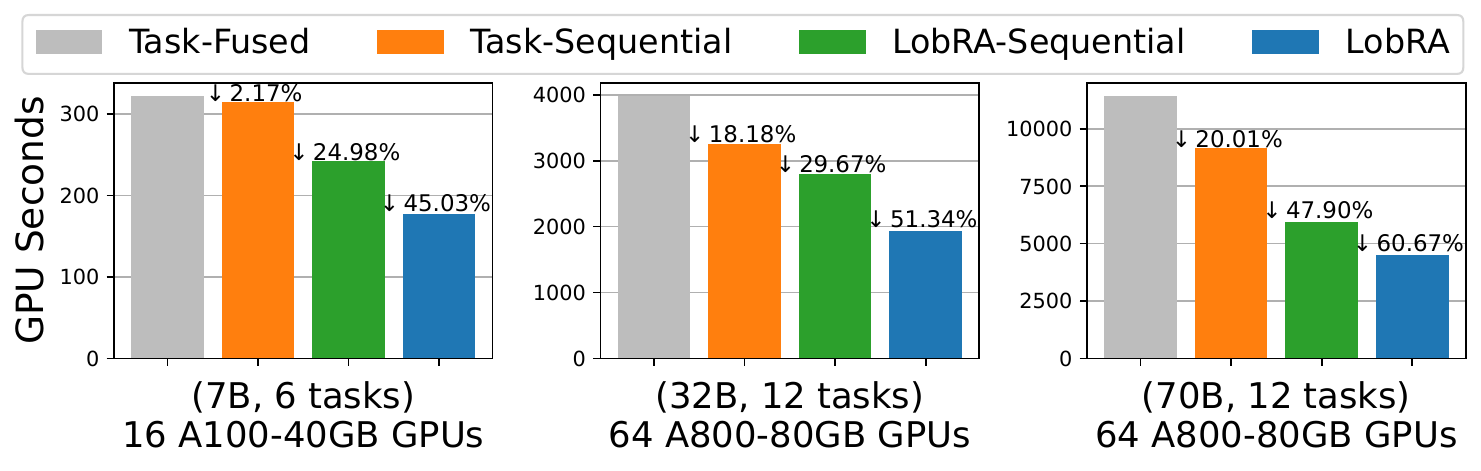}
\myvspace{-20pt}
\captionof{figure}{\small{End-to-end evaluation.}}
\label{fig:expr_e2e}
\myvspace{-10pt}
\end{figure}
\begin{table}[!t]
\centering
\captionof{table}{\small{Parallel configurations used by Task-Fused and \system in end-to-end evaluation. (Those used by Task-Sequential and \system-Sequential are provided in \ifappendix{Appendix~\ref{sec:appendix_more_expr_parallel_configs}}\else{Appendix B.3}\fi ~\cite{lobra_appendix}.) Each $\langle\alpha,\beta\rangle\times\gamma$ indicates there are $\gamma$ FT replica(s) with a TP degree of $\alpha$ and a PP degree of $\beta$.}}
\label{tb:expr_e2e_parallel_config}
\myvspace{-5pt}
\small
\begin{tabular}{|c|c|c|}
\hline
& Task-Fused & \system  \\
\hline
\hline
7B & $\langle$8,1$\rangle\times$2 & $\langle$1,1$\rangle\times$6, $\langle$2,1$\rangle\times$1, $\langle$8,1$\rangle\times$1
\\
32B & $\langle$8,1$\rangle\times$8 & $\langle$1,6$\rangle\times$4, $\langle$2,2$\rangle\times$4, $\langle$4,1$\rangle\times$2, $\langle$8,1$\rangle\times$2 
\\
70B & $\langle$16,1$\rangle\times$4 & $\langle$2,4$\rangle\times$4, $\langle$4,2$\rangle\times$1, $\langle$8,1$\rangle\times$1, $\langle$16,1$\rangle\times$1
\\
\hline
\end{tabular}
\myvspace{-10pt}
\end{table}

\section{Experiments}
\label{sec:expr}

\subsection{Implementation and Experimental Setup}
\label{sec:expr_setup}

We implement \system on top of Hetu~\cite{hetu_scis,hetu_v2}, a distributed deep learning framework for large-scale models. 
We utilize SCIP~\cite{scip} to solve the MINLP and ILP problems. 
Our framework incorporates libraries like FlashAttention~\cite{flash_attn,flash_attn_v2} for efficient computation of LLMs and NCCL~\cite{nccl} for communication. 
Even when training with homogeneous FT replicas, \system achieves state-of-the-art FT efficiency.
More details are provided in \ifappendix{Appendix~\ref{sec:appendix_nemo_compare}}\else{Appendix C}\fi ~\cite{lobra_appendix}.

Although \system focuses on co-optimizing a given batch of FT tasks, it also supports dynamic batches. 
In practice, when the batch of FT tasks changes (e.g., some tasks exit earlier than others or new FT requests arrive), we simply re-generate a new model deployment plan with the updated sequence length distribution. 
If the new plan differs from the current one, we save checkpoints for LoRA adapters and restart the joint task to meet the new plan. 
(We do not need to save checkpoints for the base model.)
The overhead of deployment adjustment is consistently less than 3 minutes in practice, which is worthwhile as the FT tasks would take hours to finish.

\textbf{Experimental Environments.}
We use two environments for evaluation. 
The first consists of 2 servers equipped with 8 A100-40GB GPUs (16 GPUs in total). 
The GPUs within the same server are connected via 600GB/s NVLink and the servers are connected via 100GB/s InfiniBand. 
The second consists of 8 servers equipped with 8 A800-80G GPUs (64 GPUs in total). 
The intra- and inter-server communication bandwidths are 400GB/s and 200GB/s, respectively. 

\textbf{Competitors.}
The primary goal of our evaluation is to evaluate the effectiveness of heterogeneous FT replicas and workload-balanced data dispatching. 
Since none of the existing works have supported such designs for joint FT, we consider two baselines, termed Task-Fused and Task-Sequential, that employ homogeneous FT replicas and uniform dispatching. 
The first na\"ively fuses the FT tasks (i.e., Figure~\ref{fig:design_anatomy}(b)) whilst the second executes the FT tasks sequentially (i.e., Figure~\ref{fig:design_anatomy}(a)). 
We tune their deployment plan to achieve the best efficiency. 
To further evaluate the effectiveness of batch co-optimization, we also consider a variant of \system, termed \system-Sequential, which also runs the FT tasks sequentially but optimizes each FT task by employing heterogeneous FT replicas and workload-balanced data dispatching.

\textbf{Workloads.}
We consider three popular LLMs, which are Llama2-7B, Qwen2.5-32B, and Llama2-70B. 
We fine-tune the 7B model with the first environment and the other two models with the second environment.
We consider 12 FT datasets, regarding each as one FT task. 
The detailed descriptions of the FT datasets and the batch size settings are provided in \ifappendix{Appendix~\ref{sec:appendix_more_expr_summary_of_task}}\else{Appendix B.1}\fi ~\cite{lobra_appendix}.
By default, we consider 6 tasks for the 7B model and 12 tasks for the other two models.
We use the Adam optimizer~\cite{adam,adamw} for all experiments.

\textbf{Protocols.}
Since our goal is to improve the efficiency of joint FT, we focus on the GPU seconds required to train one step for all involved tasks. 
For all experiments, we report the mean of 100 training steps, and the standard deviation is within 10\%.

\begin{figure}[!t]
\centering
\includegraphics[width=\linewidth]{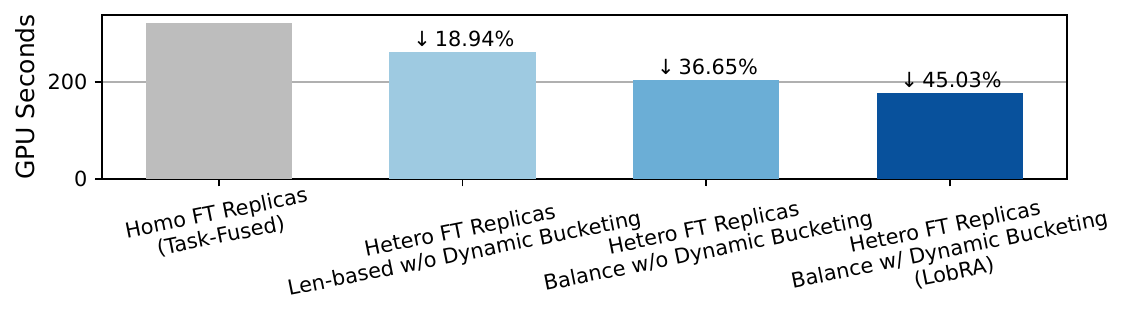}
\myvspace{-20pt}
\caption{\small{Ablation Studies (7B model, 16 A100-40GB GPUs).}}
\label{fig:expr_ablation}
\myvspace{-15pt}
\end{figure}

\begin{figure*}[!t]
\centering
\includegraphics[width=\linewidth]{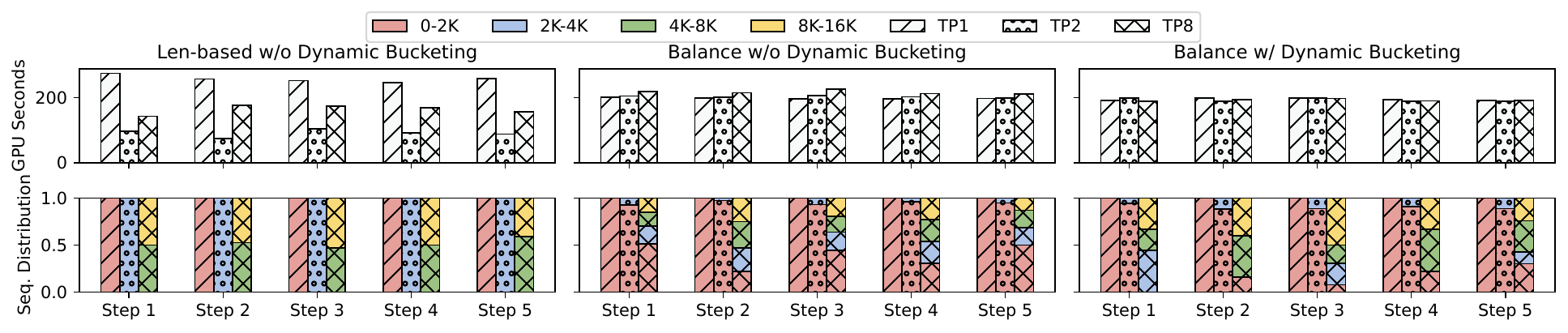}
\myvspace{-22pt}
\caption{\small{Case studies (7B model, 16 A100-40GB GPUs). Each bar represents one kind of FT replica(s). Top: The per-step time of each kind of FT replica(s). Bottom: The organization of dispatched data in terms of their sequence lengths for each kind of FT replica(s).}}
\label{fig:expr_case_study}
\myvspace{-10pt}
\end{figure*}

\subsection{End-to-End Evaluation}
\label{sec:expr_e2e}

We first assess the end-to-end joint FT efficiency of \system. 
The results are shown in Figure~\ref{fig:expr_e2e}, and we provide the parallel configurations used for deployment in Table~\ref{tb:expr_e2e_parallel_config}.

In general, \system outperforms Task-Fused significantly for all three experiments, reducing 45.03\%-60.67\% GPU seconds. 
\system achieves greater improvement when fine-tuning larger models over 64 GPUs. 
This is reasonable since \system is able to explore a larger space of deployment plans --- \system deploys 11 and 7 heterogeneous FT replicas for the 32B and 70B models, respectively, whilst Task-Fused can only deploy 8 and 4 homogeneous FT replicas. 
This allows more short sequences to get accelerated. 
Moreover, owing to the substantial size of the 70B model, Task-Fused must utilize a TP degree of 16 to accommodate the high memory consumption, which is extremely inefficient due to the slow communication across servers. 
In contrast, most of the FT replicas in \system do not need to span across servers, delivering much better efficiency. 
Thus, \system achieves the highest performance gain on the 70B model.

Task-Sequential performs better than Task-Fused. 
As analyzed in \S\ref{sec:design_anatomy}, Task-Fused must employ a high model parallel degree for all training data.
Task-Sequential, however, can employ a lower model parallel degree for datasets that contain only short sequences (e.g., question-answering datasets), so the total GPU seconds are shorter. 
The smaller performance gap in the 7B model is primarily because the smaller GPU memory capacity (40GB) restricts Task-Sequential from choosing more efficient parallel configurations for most tasks. 
Nevertheless, Task-Sequential is still less efficient than \system. 
Although \system-Sequential outperforms Task-Sequential, \system still achieves 1.32-1.44$\times$ of speedup compared to \system-Sequential. 
This is not surprising for two reasons. 
Firstly, the data heterogeneity is milder within certain tasks, so the gain of using heterogeneous FT replicas is lower. 
Secondly, the batch size for each FT task is usually small, making it difficult to achieve workload balance by routing the data across the replicas.
In practice, we find that some tasks would even experience an efficiency drop when employing \system-Sequential (detailed in \ifappendix{Appendix~\ref{sec:appendix_more_expr_results}}\else{Appendix B.2}\fi ~\cite{lobra_appendix}).
Last but not least, it is noteworthy that Task-Sequential and \system-Sequential necessitate running the FT tasks individually, either requiring extra GPUs or unfairly forcing some tasks to queue for a long time. 
Consequently, \system is more efficient and suitable for processing FT requests.

\begin{figure}[!t]
\centering
\includegraphics[width=\linewidth]{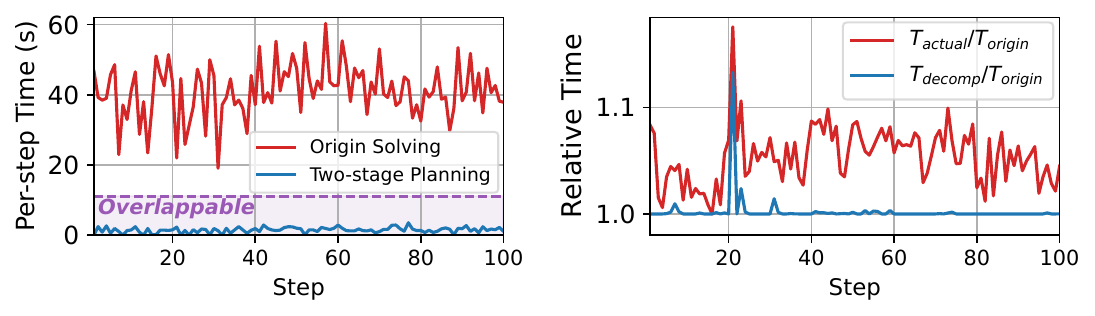}
\myvspace{-25pt}
\caption{\small{Left: Time cost of solving the original problem (Equation~\eqref{eq:origin_prob}) vs. the two-stage planning (dynamic bucketing + solving Equation~\eqref{eq:data_prob}). The horizontal dashed line indicates the average per-step time. Right: Comparison of estimated running time for solving the original problem ($T_{origin}$), the two-stage decomposition ($T_{decomp}$), and the actual running time ($T_{actual}$).}}
\label{fig:expr_effectiveness}
\myvspace{-15pt}
\end{figure}

\subsection{Effectiveness of the Proposed Techniques}
\label{sec:expr_ablation}

\mysubsubsection{Ablation and Case Studies}
We assess the effectiveness of heterogeneous model deployment and workload-balanced data dispatching, respectively. 
The results are shown in Figure~\ref{fig:expr_ablation} and Figure~\ref{fig:expr_case_study}.

When heterogeneous FT replicas are deployed and the data are directly dispatched via their lengths (i.e., Figure~\ref{fig:design_anatomy}(c)), we can reduce the GPU seconds by 18.94\% compared to the na\"ively fused approach. 
As shown in Table~\ref{tb:expr_e2e_parallel_config}, the na\"ive approach needs to employ a high model parallel degree (resulting in 2 replicas with $\langle$TP=8, PP=1$\rangle$) to support long sequences, yet it is unsuitable for short sequences. 
In contrast, with the heterogeneous model deployment, we can deploy 7 replicas with low model parallel degrees (6 replicas with $\langle$TP=1, PP=1$\rangle$ and 1 replica with $\langle$TP=2, PP=1$\rangle$). 

By further enabling the workload-balanced data dispatching and dynamic bucketing, the reduction in GPU seconds improves to 36.65\% and 45.03\%. 
Figure~\ref{fig:expr_case_study} visualizes the data dispatching and workload balance. 
When the data are simply assigned via their lengths (left-most of Figure~\ref{fig:expr_case_study}), we observe severe imbalance across the replicas due to the skewness issue. 
Workload-balanced data dispatching (middle of Figure~\ref{fig:expr_case_study}) strikes a good balance among the replicas by routing the data. 
With dynamic bucketing (right-most of Figure~\ref{fig:expr_case_study}), the running time can be further reduced, particularly for replicas with a higher model parallel degree, since dynamic bucketing is more effective in reducing the padding of longer sequences.

\mysubsubsection{Effectiveness of Planning}
As discussed in \S\ref{sec:method_overview}, we employ a two-stage decomposition to Equation~\eqref{eq:origin_prob}. 
One reason is that its solving is time-consuming. 
To examine this, we measure its solving time and compare it with the per-step running time. As shown on the left side of Figure~\ref{fig:expr_effectiveness}, solving Equation~\eqref{eq:origin_prob} is slower than one training step, even if we have applied the configuration pruning heuristics introduced in \S\ref{sec:method_deployment} (the effectiveness of the two heuristics is evaluated in \ifappendix{Appendix~\ref{sec:appendix_more_expr_results}}\else{Appendix B.2}\fi ~\cite{lobra_appendix}). 
On the contrary, with the two-stage decomposition, we only need to perform the dynamic bucketing and solve Equation~\eqref{eq:data_prob} for each step, which is extremely efficient and can be fully overlapped by the training of previous step(s).

In addition, we assess how the two-stage decomposition impacts the efficiency of the achieved solutions. 
To do so, we record the estimated running time given by solving the original problem, the estimated running time after the two-stage decomposition, and the actual running time in practice, denoted as $T_{origin}, T_{decomp}, T_{actual}$, respectively.
The right side of Figure~\ref{fig:expr_effectiveness} presents $T_{decomp}/T_{origin}$ and $T_{actual}/T_{origin}$ across 100 steps. 
Overall, $T_{decomp}$ and $T_{origin}$ are extremely close in most steps. 
In occasional steps, the sampled batch does not contain long sequences, so solving the original problem produces a better deployment plan, leading to several spikes. 
However, the performance gap is still small (within 15\%). 
These results verify the robustness and effectiveness of our two-stage decomposition. 
Besides, $T_{actual}$ is also close to $T_{decomp}$ (within 10\%) across all steps, demonstrating the accuracy of our cost model.

\mysubsubsection{More Experiments}
We have conducted more experiments, including the scalability w.r.t. the number of GPUs and tasks, the sensitivity w.r.t. the sequence bucketing, and the effectiveness of our configuration pruning. 
Due to the space constraint, we leave more results and details of our experiments in \ifappendix{Appendix~\ref{sec:appendix_more_expr}}\else{Appendix B}\fi ~\cite{lobra_appendix}.

\section{Conclusion}
\label{sec:conc}

This work studies the processing of multiple FT requests by jointly training multiple LoRA adapters with the same base model. 
We conducted an anatomy on the efficiency of joint FT, showing that two data heterogeneity issues, i.e., the sequence length variation and skewness, pose significant challenges. 
Then, we developed a brand new joint-FT framework, namely \system, with two innovative designs. 
The first is the deployment of heterogeneous FT replicas, which addresses the sequence length variation and accelerate the processing of short sequences. 
The second is workload-balanced data dispatching, which eliminates the idle period of some FT replicas caused by skewness. 
Empirical results show that \system greatly reduces the GPU seconds required for joint FT by 45.03\%-60.67\%.

\begin{acks}
This work is supported by National Science and Technology Major Project (2022ZD0116315), National Natural Science Foundation of China (U23B2048, 62402011), Beijing Municipal Science and Technology Project (Z231100010323002), ZTE-PKU joint program, and High-performance Computing Platform of Peking University. 
Fangcheng Fu and Bin Cui are the corresponding authors.
\end{acks}

\balance
\bibliographystyle{ACM-Reference-Format}
\bibliography{references}

\ifappendix

\onecolumn
\clearpage
\appendix

\section{Details of Configuration Pruning}
\label{sec:appendix_config_pruning}

As introduced in \S\ref{sec:method_deployment}, in order to accelerate the problem-solving of model deployment (i.e., Equation~\eqref{eq:model_prob}), we propose the configuration pruning heuristics. 
Due to the space constraint, we only present the rationale of the two heuristics. 
In this section, we provide details about how to propose candidate parallel configurations and estimate the efficiency lower bound of a deployment plan.

\mysubsubsection{Configuration Proposal}
The first is to propose a small set of parallel configurations (rather than covering all possible ones) for problem-solving, which is based on the fact that many parallel configurations are consistently less efficient than others. 
To elaborate, we conduct an empirical study and find that a partial order relation on the efficiency exists among the configurations, as described in Observation \ref{theo:appendix_partial_order}.

\begin{observation}
\label{theo:appendix_partial_order}
Given a sequence length $s_0$ and two parallel configurations $\mathcal{S}_{\alpha}$ and $\mathcal{S}_{\beta}$, if $\mathcal{S}_{\alpha}$ has higher throughput (i.e., processed tokens per GPU per second) on sequence length $s_0$, then for any sequence length $s < s_0$, if the batch size $b$ satisfies $b \times s = s_0$, $\mathcal{S}_{\alpha}$ has higher throughput than $\mathcal{S}_{\beta}$.
\end{observation}

To validate the observation presented above, we conduct a series of experiments on different configurations. As shown in Table~\ref{tb:parallel_config_thruputs}, for each configuration, we evaluate its throughput when faced with different sequence lengths. Taking $s_0=8K$ and \textit{num\_gpus} = 8 as an example, configuration $\langle$TP=2, PP=4$\rangle$ achieves higher throughput than other configurations with the same sequence length, and its throughput is still higher for shorter sequences, which matches Observation \ref{theo:appendix_partial_order}.

Based on this, when faced with a large number of parallel configurations, we propose candidate configurations by selecting those with the highest throughput for each (\textit{num\_gpus}, \textit{seq\_len}) pair, which can be depicted by the following SQL: 
\begin{center}
\texttt{SELECT} \textit{config}, \texttt{MAX}(\textit{thruput}) \texttt{FROM} \textit{thruput\_table} \texttt{GROUP BY} \textit{num\_gpus, seq\_len}    
\end{center}
By doing so, a configuration will not be selected if it is consistently outperformed by the others, so the achieved solutions will not be affected (as evaluated in Appendix~\ref{sec:appendix_more_expr_results}). 
In addition, it is obvious that the number of candidate configurations is limited to at most $O(R\log{N})$, where $R$ and $N$ denote the number of buckets and GPUs, respectively. 
Consequently, we can effectively reduce the number of candidate configurations, whilst guaranteeing correctness.

\begin{table}[H]
\centering
\caption{\small{Throughput (i.e., processed tokens per GPU per second) of each candidate parallel configuration with sequence lengths and different numbers of GPUs. A cell is left empty if there are not enough GPUs to deploy with the corresponding parallel configuration. The symbol ``\ding{55}'' indicates the parallel configuration does not support processing the corresponding sequence length due to out-of-memory errors. The symbol ``-'' indicates the throughput remains the same after model replication.}}
\label{tb:parallel_config_thruputs}
\begin{tabular}{|c|cccc|ccc|cc|c|}
\hline
\textit{seq\_len} & \multicolumn{4}{c|}{2K} & \multicolumn{3}{c|}{4K} & \multicolumn{2}{c|}{8K} & \multicolumn{1}{c|}{16K} \\
\hline
\textit{num\_gpus} & 1 & 2 & 4 & 8 & 2 & 4 & 8 & 4 & 8 & 8
\\
\hline
\hline
$\langle$TP=1, PP=1$\rangle$ 
& \textit{\textbf{5.11}} & - & - & - 
& \ding{55} & \ding{55} & \ding{55} 
& \ding{55} & \ding{55} 
& \ding{55} 
\\
\hline
$\langle$TP=2, PP=1$\rangle$ 
& & 4.30 & - & - 
& \textit{\textbf{4.12}} & - & -
& \ding{55} & \ding{55} 
& \ding{55} 
\\
$\langle$TP=1, PP=2$\rangle$ 
& & 4.88 & - & -
& \ding{55} & \ding{55} & \ding{55} 
& \ding{55} & \ding{55} 
& \ding{55} 
\\
\hline
$\langle$TP=4, PP=1$\rangle$ 
& & \ding{55} & 3.63 & -
& & 3.50 & -
& \textit{\textbf{3.25}} & -
& \ding{55} 
\\
$\langle$TP=2, PP=2$\rangle$ 
& & \ding{55} & 4.15 & -
& & 3.98 & -
& \ding{55} & \ding{55} 
& \ding{55} 
\\
$\langle$TP=1, PP=4$\rangle$ 
& & \ding{55} & 5.03 & -
& & \textit{\textbf{4.78}} & -
& \ding{55} & \ding{55} 
& \ding{55} 
\\
\hline
$\langle$TP=8, PP=1$\rangle$ 
& & & & 2.79
& & & 2.71
& & 2.56 
& \textit{\textbf{2.33}}
\\
$\langle$TP=4, PP=2$\rangle$ 
& & & & 3.48
& & & 3.34 
& & 3.12 
& \ding{55}
\\
$\langle$TP=2, PP=4$\rangle$ 
& & & & 4.27
& & & 4.10 
& & \textit{\textbf{3.79}} 
& \ding{55}
\\
$\langle$TP=1, PP=8$\rangle$ 
& & & & 4.45
& & & 4.25 
& & \ding{55} 
& \ding{55}
\\
\hline
\end{tabular}
\end{table}

\mysubsubsection{Lower Bound Filtering} 
Secondly, we take lower bound estimation as a guideline to filter out inefficient deployment plans, leveraging length-based data dispatching and the rationale of data migration (or movement). 
In fact, considering the negative correlation between sequence length and throughput of a parallel configuration (i.e., lower throughput for longer sequences, which holds in practice), length-based data dispatching serves as a greedy approach --- with length-based data dispatching, data is always assigned to the most efficient configuration that suffices the memory consumption requirement. 
However, as discussed earlier regarding the sequence length skewness in \S\ref{sec:design_anatomy}, workload balancing necessitates re-dispatching data, which may result in some short sequences being assigned to less efficient configurations compared to length-based data dispatching. Thus, length-based data dispatching can serve as an effective method for estimating the lower bound of running time. Specifically, it is formalized as the following theorem.

\begin{theorem}
\label{theo:appendix_lower_bound}
Consider a deployment plan with $n$ heterogeneous FT replicas, each with a different maximum supportable sequence length. The corresponding number of GPUs is given by $\{N_1, N_2, \cdots, N_n\}$, and the running times when applying the length-based data dispatching are denoted by $\{t_1, t_2, \cdots, t_n\}$. After re-dispatching the training data according to the workload balancing principle, the running times adjust to $\{t_1', t_2', \cdots, t_n'\}$. Then, we have the following inequality:
\begin{equation}
\label{eq:appendix_gpu_second}
N\hat{t} \ge \sum_{i=1}^{n}{N_i t_i}
\end{equation}
where $N$ is the total number of GPUs and $\hat{t}$ denotes the maximum running time across the $n$ replicas after workload-balancing, denoted as $\hat{t} = \max_{1 \le i \le n}\{t_i'\}$.
\end{theorem}

The proof of this theorem relies on the following definitions, properties and assumptions:

\begin{definition}
\label{def:appendix_lower_bound}

The \textit{Average Throughput Bound} (\textit{ATB}) for a parallel configuration with $N$ GPUs is defined as follows:

Under the constraint that the number of chunks (i.e. micro-batches) is sufficiently large, for a given sequence length $s$ and maximum supportable sequence length $M$, let the batch size $b$ satisfies $b \times s \le M < (b + 1) \times s$, and the running time for an input size of $b \times s$ is $t$, then the \textit{ATB} for $s$ is defined as
\begin{equation}
    \textit{ATB}_{s} = \frac{b s}{N t}
\end{equation}
\end{definition}

\begin{property}
\label{prpt:appendix_lower_bound_atb}

Based on Observation \ref{theo:appendix_partial_order}, for a given sequence length $s_0$, if configuration $\mathcal{S}_{\alpha}$ and $\mathcal{S}_{\beta}$ satisfies $\textit{ATB}_{s_0, \alpha} \ge \textit{ATB}_{s_0, \beta}$, then for any sequence length $s \le \min\{M_\alpha, M_\beta\}$, it holds that $\textit{ATB}_{s,\alpha} \ge \textit{ATB}_{s, \beta}$. This property is denoted as $\textit{ATB}_{\alpha} \ge \textit{ATB}_{\beta}$.
\end{property}

\begin{property}
\label{prpt:appendix_lower_bound_pp}

Suppose $n$ replicas are arranged in descending order of \textit{ATB}. Then, the maximum supportable length of each replica increases monotonically. Consequently, during data migration for workload balancing, the $i$-th replica can only receive data from the first $i-1$ replicas.
\end{property}

\begin{assumption}
\label{aspt:appendix_lower_bound}
The relationship between the ATB of configuration $\mathcal{S}_{\alpha}$ and $\mathcal{S}_{\beta}$ can reflect the corresponding average throughput per GPU relationship during runtime, under the constraint that the number of chunks is sufficiently large. Let the average throughput per GPU during the runtime of configuration $\mathcal{S}_{\alpha}$ and $\mathcal{S}_{\beta}$ be $\overline{T}_\alpha$ and $\overline{T}_\beta$, if $\textit{ATB}_\alpha \ge \textit{ATB}_\beta$, then $\overline{T}_\alpha \ge \overline{T}_\beta$.
\end{assumption}

\begin{proof}

    The proof for Theorem \ref{theo:appendix_lower_bound} proceeds by mathematical induction on the number of replicas, denoted by $n$. In the proof, the replicas are ordered in descending order of \textit{ATB}.
    \begin{enumerate}
        \item \textbf{Base Case.} When $n = 2$, let the number of GPUs for the two replicas be $N_1$ and $N_2$, and the number of input tokens for each replica before workload balancing be $M_1$ and $M_2$, respectively. During workload balancing, $M$ tokens are moved from replica 1 to replica 2. By Assumption \ref{aspt:appendix_lower_bound}, we have
        \begin{equation}
            \begin{aligned}
                (N_1t_1' + N_2t_2') - (N_1t_1 + N_2t_2) &= (\frac{N_1(M_1 - M)}{T_1} + \frac{N_2(M_2+M)}{T_2}) - (\frac{N_1M_1}{T_1} + \frac{N_2M_2}{T_2}) \\
                &= M \times (\frac{N_2}{T_2} - \frac{N_1}{T_1}) \\
                &= M \times (\frac{1}{\overline{T}_2} - \frac{1}{\overline{T}_1}) \\
                &\ge 0
            \end{aligned}
        \end{equation}
        Thus, we obtain the inequality $N_1t_1 + N_2t_2 \le N_1t_1' + N_2t_2' \le (N_1 + N_2) \hat{t} = N\hat{t}$.
        \item \textbf{Inductive Hypothesis.} Assume that for $n \le k$, the inequality $N \hat{t} \ge \sum_{i=1}^{n}{N_i t_i}$ holds.
        \item \textbf{Inductive Step.} When $n = k + 1$, let $i$ be the smallest index such that for all $1 \le j < i$, no data are moved to other replicas during data migration for workload balancing. We discuss this in two cases:
        \begin{enumerate}
            \item If $i > 1$, by Property \ref{prpt:appendix_lower_bound_pp}, the running time of the first $i - 1$ replicas remains unchanged during workload balancing, so $t_j = t_j'$ for all $1 \le j < i$. For the remaining $n - i + 1$ replicas, by the inductive hypothesis, we have $\sum_{j=i}^{k+1}{N_j t_j} \le \sum_{j=i}^{k+1}{N_j t_j'}$. Therefore, we have
            \begin{equation}
                \sum_{j=1}^{k+1}{N_j t_j} \le \sum_{j=1}^{i-1}{N_j t_j} + \sum_{r=i}^{k+1}{N_r t_r'} = \sum_{j=1}^{k+1} {N_j t_j'}
            \end{equation}
            \item If $i = 1$, during workload balancing, replica 1 migrates sequences to other replicas. Let $M$ denote the set of data moved from replica 1. The migration process can be broken down into two steps:
            \begin{enumerate}
                \item Migrate all of $M$ to replica 2.
                \item Replica 2 then re-dispatch $M$ to other replicas to achieve workload balancing.
            \end{enumerate}
            After the first step, the running time of replica 1 and 2 are $t_1'$ and $\tilde{t}_2$, respectively (replica 1 does not receive data from other replicas, so it has achieved workload balancing). By the base case of the induction, we have $N_1 t_1 + N_2 t_2 \le N_1 t_1' + N_2 \tilde{t}_2$. By the inductive hypothesis, for the remaining $k$ replicas excluding replica 1, we have $N_2 \tilde{t}_2 + \sum_{i=3}^{k+1}{N_i t_i} \le \sum_{i=2}^{k+1}{N_i t_i'}$. Therefore, we obtain
            \begin{equation}
                \sum_{i=1}^{k+1}{N_i t_i} \le N_1 t_1' + (N_2 \tilde{t}_2 + \sum_{j=3}^{k+1}{N_j t_j}) \le N_1t_1' + \sum_{j=2}^{k+1}{N_j t_j'} = \sum_{i=1}^{k+1}{N_i t_i'}
            \end{equation}
        \end{enumerate}
    \end{enumerate}
    By induction, the conclusion holds.
\end{proof}

In the scenario where two replicas share the same maximum supportable sequence length, we can group these replicas together, and perform workload balancing within the group. This allows each group to be treated as a single replica in the proof process, thereby reducing the problem to the case where the maximum supportable sequence lengths of all replicas are distinct.

Theorem \ref{theo:appendix_lower_bound} suggests an efficient and effective method for estimating the lower bound of a deployment plan, expressed as $\frac{\sum_{i=1}^{n}{N_i t_i}}{N}$. The estimation leverages length-based data dispatching, which relies solely on sequence lengths and does not account for the workload relationships among different replicas. 

However, Assumption \ref{aspt:appendix_lower_bound} does not always hold in real-world scenarios where the number of chunks is variable and may be insufficient. As a result, in this work, we only treat the lower bound as a relative metric to filter out deployment plans whose estimated lower bounds exceed the current minimum by more than a pre-defined threshold ($15\%$ by default). This approach is demonstrated to be effective in practice.

\mysubsubsection{Putting them together}
In conclusion, we summarize the configuration pruning process as the following steps:

\begin{enumerate}
    \item Propose configuration candidates leveraging the offline benchmarking results.
    \item Construct possible deployment plans using the parallel configuration candidates, formulated as an integer partition problem, and solved via dynamic programming.
    \item Estimate the lower bound for each deployment plan and filter out those that are estimated to be inefficient. 
\end{enumerate}

Eventually, for each remaining deployment plan, we determine its performance individually by solving the problem defined in Equation~\ref{eq:model_prob}. It is noteworthy that, if a deployment plan is given, then the problem turns into an ILP problem, which is much more efficient to solve. 
Thus, we enumerate all remaining deployment plans, solve the problem for each one (in parallel), and select the most efficient one as the final deployment plan.

In Appendix~\ref{sec:appendix_more_expr_results}, we will provide experiments to show that the solving of the model deployment plan can be accelerated significantly, whilst maintaining the same solution.

\clearpage

\section{More Experimental Details and Results}
\label{sec:appendix_more_expr}

\subsection{Summary of Evaluated FT Tasks}
\label{sec:appendix_more_expr_summary_of_task}

We select 12 popular open-source FT datasets and treat each as one FT task, spanning instruction tuning, question answering, and summarization tasks. These datasets encompass diverse fields, including mathematics, coding, medicine, and routine task handling. The summary is presented in Table \ref{tb:dataset_summary}.

\begin{table}[H]
\centering
\caption{\small{Summary of FT datasets used in our experiments.}}
\label{tb:dataset_summary}
\begin{tabular}{|c|c|c|c|c|c|}
\hline
\textbf{Dataset Name} & \textbf{\begin{tabular}[c]{@{}c@{}}Avg. Sequence\\ Length\end{tabular}} & \textbf{Skewness} & \textbf{Kurtosis} & \textbf{Type of Task} & \textbf{Batch Size} \\ 
\hline \hline
databricks-dolly-15k\tablefootnote{https://huggingface.co/datasets/databricks/databricks-dolly-15k} & 207 & 7.11 & 95.43 & Instruction Tuning & 256 \\
\hline
python\_code\_instructions\tablefootnote{https://huggingface.co/datasets/iamtarun/python\_code\_instructions\_18k\_alpaca} & 269 & 10.01 & 121.55 & Code Instruction Tuning & 128 \\
\hline
Evol-Instruct\tablefootnote{https://huggingface.co/datasets/ise-uiuc/Magicoder-Evol-Instruct-110K} & 702 & 6.59 & 80.28 & Code Instruction Tuning & 128 \\
\hline
CommitPackFt\tablefootnote{https://huggingface.co/datasets/chargoddard/commitpack-ft-instruct} & 663 & 0.79 & 1.68 & Code Instruction Tuning & 128 \\
\hline
MathInstruct\tablefootnote{https://huggingface.co/datasets/TIGER-Lab/MathInstruct} & 252 & 3.03 & 12.72 & Math Instruction Tuning & 128 \\
\hline
MetaMathQA\tablefootnote{https://huggingface.co/datasets/meta-math/MetaMathQA} & 236 & 2.56 & 14.56 & Math Question Answering & 128 \\
\hline
NuminaMath-CoT\tablefootnote{https://huggingface.co/datasets/AI-MO/NuminaMath-CoT} & 543 & 1.52 & 3.51 & Math Question Answering & 256 \\
\hline
PubMedQA\tablefootnote{https://huggingface.co/datasets/qiaojin/PubMedQA} & 371 & 0.73 & 3.29 & Medical Question Answering & 64 \\
\hline
XSum\tablefootnote{https://huggingface.co/datasets/EdinburghNLP/xsum} & 526 & 7.49 & 371.80 & Summarization & 128 \\
\hline
BillSum\tablefootnote{https://huggingface.co/datasets/FiscalNote/billsum} & 3903 & 0.85 & 0.30 & Summarization & 32 \\
\hline
cnn\_dailymail\tablefootnote{https://huggingface.co/datasets/abisee/cnn\_dailymail} & 947 & 0.89 & 0.64 & Summarization & 256 \\
\hline
MeetingBank\tablefootnote{https://huggingface.co/datasets/huuuyeah/meetingbank} & 3622 & 4.35 & 26.50 & Summarization & 64 \\ 
\hline
\end{tabular}
\end{table}

For the instruction tuning tasks, the databricks-dolly-15k dataset is a corpus of approximately 15,000 records generated by humans, covering various instruction categories, such as information extraction. The python\_code\_instructions dataset contains more than 18,600 problem descriptions and code in Python language, formatted in Alpaca style. The Evol-Instruct dataset augments the code instruction tuning dataset CodeAlpaca\_20K via a total of 10 augmentation strategies, covering more than 11,000 samples. The CommitPackFT dataset leverages the natural structure of Git commits for instruction tuning, comprising about 491,000 pairs of code changes and human instructions and covering 4 terabytes of Git commits across 350 programming languages. The MathInstruct dataset is compiled from 13 math rationale datasets and focuses on the hybrid use of rationales like chain-of-thought, containing over 262,000 mathematical problem-solution pairs.

For the question-answering tasks, the MetaMathQA dataset augments the training sets of the GSM8K and MATH datasets by rewriting the questions from multiple perspectives, comprising roughly 395,000 mathematical question-answering samples. The NuminaMath-CoT dataset consists of approximately 860,000 mathematical problems, where each solution is formatted in a chain-of-thought manner. The PubMedQA dataset is a dataset for biomedical research question answering, and it has 211,300 artificially generated QA instances.

For the summarization tasks, the XSum dataset consists of approximately 204,000 text-summary pairs from the BBC news dataset. The BillSum dataset is a corpus of summarization of US Congressional and California state bills, compromising 189,00 text-summary pairs. The cnn\_dailymail dataset contains over 300,000 unique news articles written by journalists at CNN and the Daily Mail and abstracts. The MeetingBank dataset contains 6,892 segment-level summarization instances from council meetings.

\subsection{More Experimental Results}
\label{sec:appendix_more_expr_results}

Below we present more experimental results to thoroughly evaluate the effectiveness of our work.

\mysubsubsection{Scalability}
We evaluate the scalability w.r.t. the number of GPUs and tasks, respectively. The results are provided in Figure~\ref{fig:expr_scalability}.
We first compare the GPU seconds of 4-task joint FT over varying numbers of GPUs. 
With only 16 GPUs, both \system and Task-Fused can only deploy one FT replica (with $\langle$TP=16,PP=1$\rangle$), so they have the same performance. 
However, with more GPUs, \system can leverage heterogeneous FT replicas, reducing the GPU seconds needed for each training step. 
On the contrary, Task-Fused's efficiency goes down slightly when there are more GPUs, owing to the increased overhead of synchronization among FT replicas.  
Then we measure how the GPU seconds change w.r.t. the number of tasks using the 70B model over 64 GPUs. 
It can be observed that both \system and Task-Fused exhibit a near-linear increment in the GPU seconds when there are more tasks, which is reasonable as the FT workloads become higher. 
Yet \system consistently outperforms Task-Fused with all numbers of tasks.

\mysubsubsection{Sensitivity}
To process training data with different levels of lengths using divergent parallel configurations, our work introduces a hyper-parameter $R$, which indicates that we divide the training data into $R$ buckets according to their lengths. 
To assess the sensitivity of our work w.r.t. $R$, we enumerate the value of $R$ from 4 to 32 and record the training time as well as how many padding tokens are added.
The results are shown in Figure~\ref{fig:expr_sensitivity}. 
It can be observed that, as $R$ increases, it consistently reduces the padding tokens, which is reasonable. 
However, after $R$ goes beyond 12, the training time remains stable despite the reduction in padding tokens, which is because more buckets would lead to more overhead. 
In short, the efficiency of \system is robust to the choice of $R$.

\begin{figure}[H]
\begin{minipage}{0.59\linewidth}
\centering
\includegraphics[width=\linewidth]{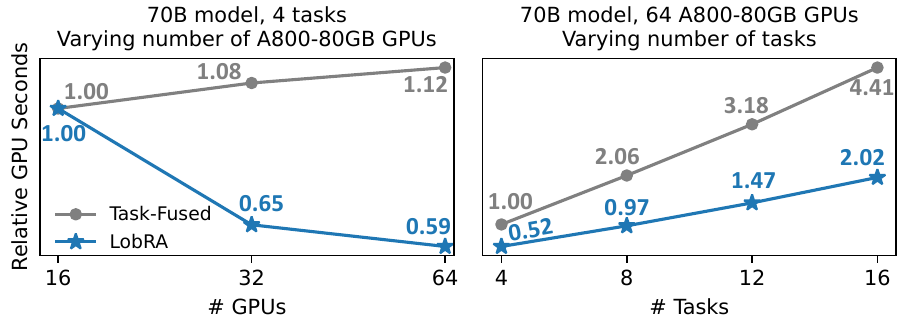}
\captionof{figure}{\small{Scalability w.r.t. numbers of GPUs and tasks (70B model).}}
\label{fig:expr_scalability}
\hfill
\end{minipage}
\begin{minipage}{0.39\linewidth}
\centering
\includegraphics[width=\linewidth]{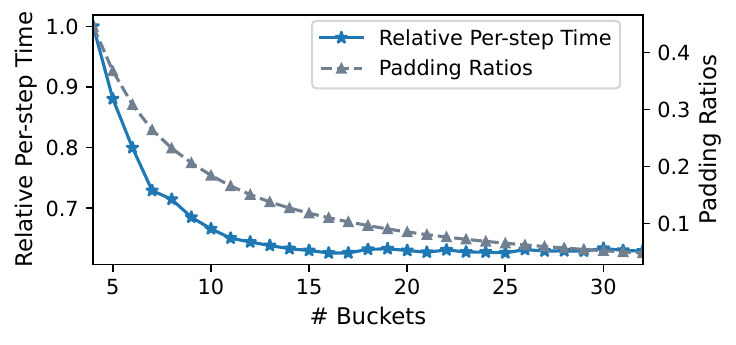}
\caption{\small{Impact of number of buckets (i.e., $R$) in dynamic bucketing to the per-step time and padding ratios (7B model, 16 A100-40GB GPUs). The per-step time is scaled by that with 4 buckets.}}
\label{fig:expr_sensitivity}
\end{minipage}
\end{figure}

\mysubsubsection{Effectiveness of Configuration Planning}
To assess the scalability of configuration planning (i.e., solving Equation~\eqref{eq:model_prob}) as well as the effectiveness of the two configuration pruning techniques, we conduct experiments to measure the time cost of solving Equation~\eqref{eq:model_prob} for the 70B model using 16 to 256 GPUs.
The datasets are consistent with those used in the GPU scalability evaluation, detailed in Appendix \ref{sec:appendix_more_expr_parallel_configs}.
The results are presented in Table \ref{tb:appendix_config_pruning}. 
Specifically, the parallel configurations solved by all approaches are the same (as long as there is no timeout), validating the accuracy of our configuration pruning methods. 
Furthermore, although the first pruning technique (i.e., configuration proposal) accelerates the solving process, it still encounters timeout issues when the number of GPUs exceeds 48. 
In contrast, enabling both pruning techniques (i.e., configuration proposal and lower bound filtering) efficiently delivers results within several minutes, even in the case of 256 GPUs, demonstrating its scalability.
Although there is still an overhead of solving the model deployment at initialization, it is worthwhile given the speedup in the joint FT process. 
In addition, for FT, it is rare to use 256 GPUs or more, so our work fits real-world FT scenarios well.

\begin{table}[H]
\centering
\caption{\small{Time cost (in seconds) of configuration planning via different approaches (70B model). The symbol ``{\ding{55}}'' indicates that the problem solving fails to finish within 3600 seconds (1 hour). The achieved deployment plan is consistent across all approaches.}}
\label{tb:appendix_config_pruning}
\small
\begin{tabular}{|c|c|c|c|c|}
\hline
\textbf{\# GPUs} & \textbf{\begin{tabular}[c]{@{}c@{}}w/o Configuration Proposal\\ w/o Lower Bound Filtering\end{tabular}} & \textbf{\begin{tabular}[c]{@{}c@{}}w/ Configuration Proposal\\ w/o Lower Bound Filtering\end{tabular}} & \textbf{\begin{tabular}[c]{@{}c@{}}w/ Configuration Proposal\\ w/ Lower Bound Filtering\end{tabular}} & \textbf{Deployment Plan} \\
\hline \hline
16 & 0.067 & 0.034 & 0.014 &  $\langle$16,1$\rangle\times$1 \\ \hline
24 & 0.842 & 0.540 & 0.378 &  $\langle$2,4$\rangle\times$1, $\langle$16,1$\rangle\times$1 \\ \hline
32 & 7.63 & 3.29 & 1.71 & $\langle$2,4$\rangle\times$1, $\langle$4,2$\rangle\times$1, $\langle$16,1$\rangle\times$1 \\ \hline
40 & \ding{55} & 264.24 & 28.78 & $\langle$2,4$\rangle\times$1, $\langle$4,2$\rangle\times$1, $\langle$8,1$\rangle\times$1, $\langle$16,1$\rangle\times$1 \\ \hline
48 & \ding{55} & 2957.01 & 48.84 & $\langle$2,4$\rangle\times$2, $\langle$4,2$\rangle\times$1, $\langle$8,1$\rangle\times$1, $\langle$16,1$\rangle\times$1 \\ \hline
64 & \ding{55} & \ding{55} & 148.81 & $\langle$2,4$\rangle\times$3, $\langle$4,2$\rangle\times$1, $\langle$8,1$\rangle\times$2, $\langle$16,1$\rangle\times$1 \\ \hline
128 & \ding{55} & \ding{55} & 239.84 & $\langle$2,4$\rangle\times$5, $\langle$8,1$\rangle\times$9, $\langle$16,1$\rangle\times$1 \\ \hline
256 & \ding{55} & \ding{55} & 428.08 & $\langle$2,4$\rangle\times$3, $\langle$4,1$\rangle\times$16, $\langle$4,2$\rangle\times$6, $\langle$8,1$\rangle\times$13, $\langle$16,1$\rangle\times$1 \\ \hline
\end{tabular}
\end{table}

\mysubsubsection{Comparison between Task-Sequential and \system-Sequential}
To evaluate the effectiveness of our designs in single-task FT, we compare the training time of each task of Task-Sequential and \system-Sequential. 
It can be seen that although \system-Sequential achieves over 60\% speedup on 1 task (which even surpasses the end-to-end speedup achieved by \system), the speedup on the other tasks is lower, and it even suffers from performance degradation on 2 tasks. 
This phenomenon is reasonable. 
For one thing, employing heterogeneous FT replicas is less beneficial for certain tasks whose data heterogeneity is not severe.
For another, unlike pre-training, the batch size for each FT task is usually small in order to avoid over-fitting, as a result of which, it is difficult to achieve workload balance by adjusting the data dispatching across replicas.

\begin{table}[H]
\centering
\caption{\small{Comparison of GPU seconds for each FT task of Task-Sequential and \system-Sequential (70B model).}}
\label{tb:appendix_compare_two_sequential_approaches}
\begin{tabular}{|c|c|c|c|}
\hline
\textbf{Dataset Name} & \textbf{Task-Sequential} ($T_1$) & \textbf{\system-Sequential} ($T_2$) & $(T_1 - T_2) / T_1$ \\
\hline \hline
MathInstruct & 209.5808 &	162.7456	& 22.35\% \\ \hline
python\_code\_instructions &	539.232 &	296.3648 &	45.04\% \\ \hline
databricks-dolly-15k &	404.4928 &	264.7744 &	34.54\% \\ \hline
BillSum &	1231.0016 &	800.7936 &	34.95\%  \\ \hline
CommitPackFt &	471.6864 &	278.7072 &	40.91\% \\ \hline
NuminaMath-CoT &	697.8368 &	471.9232 &	32.37\% \\ \hline
PubMedQA &	129.9712 &	157.6768 &	-21.32\% \\ \hline
MetaMathQA &	168.1856 &	117.7536 &	29.99\% \\ \hline
Evol-Instruct &	1250.3296 &	476.224 &	61.91\% \\ \hline
cnn\_dailymail &	1074.112 &	1227.7312 &	-14.30\% \\ \hline
XSum &	786.4 &	417.632 &	46.89\% \\ \hline
MeetingBank &	2186.4192 &	1249.5936 &	42.85\% \\ \hline
\end{tabular}
\end{table}

\subsection{Details of Experiment Configurations}
\label{sec:appendix_more_expr_parallel_configs}

In this section, we provide more details about the datasets and parallel configurations used in our experiments.

For the end-to-end evaluation in \S\ref{sec:expr_e2e}, we consider all 12 tasks listed in Table \ref{tb:dataset_summary} for the 32B and 70B models. For the 7B model, we focus on a subset of 6 tasks: databricks-dolly-15k, Evol-Instruct, XSum, CommitPackFT, MeetingBank, and python\_code\_instructions. Furthermore, the specific parallel configurations of various tasks in the Task-Sequential and \system-Sequential baselines are presented in Table \ref{tb:appendix_task_sequential_config} and Table \ref{tb:appendix_lobra_sequential_config}, respectively.
Besides, the experiments in \S\ref{sec:expr_ablation} use the same parallel configurations as those in the end-to-end evaluation.

For the GPU and task scalability evaluation in Figure \ref{fig:expr_scalability}, we focus on a subset of 4 tasks: Evol-Instruct, CommitPackFT, BillSum, and PubMedQA. Table \ref{tb:appendix_gpu_scalability_config} and Table \ref{tb:appendix_task_scalability_config} present the parallel configurations of Task-Fused and \system under varying GPU counts and task numbers.

\begin{table}[H]
\centering
\captionof{table}{\small{Parallel configurations used by Task-Sequential for the end-to-end evaluation. (Those used by Task-Fused and \system are provided in Table~\ref{tb:expr_e2e_parallel_config}.) Each $\langle\alpha,\beta\rangle\times\gamma$ indicates that there are $\gamma$ FT replica(s) with a TP degree of $\alpha$ and a PP degree of $\beta$.}}

\label{tb:appendix_task_sequential_config}
\begin{tabular}{|c||c|c|c|}
\hline
\multirow{2}*{\textbf{Dataset Name}} & \multicolumn{3}{c|}{\textbf{Task-Sequential}} \\
\cline{2-4}
& \textbf{7B} & \textbf{32B} & \textbf{70B} \\ \hline 
\hline
MathInstruct 
& - & $\langle$2,2$\rangle\times$16 & $\langle$4,2$\rangle\times$8 \\
\hline
python\_code\_instructions
& $\langle$8,1$\rangle\times$2 & $\langle$8,1$\rangle\times$8 & $\langle$16,1$\rangle\times$4 \\
\hline
databricks-dolly-15k
& $\langle$4,1$\rangle\times$4 & $\langle$4,1$\rangle\times$16 & $\langle$8,1$\rangle\times$8 \\ 
\hline
BillSum
& - & $\langle$8,1$\rangle\times$8 & $\langle$16,1$\rangle\times$4 \\
\hline
CommitPackFt
& $\langle$4,1$\rangle\times$4 & $\langle$4,1$\rangle\times$16 & $\langle$8,1$\rangle\times$8 \\ 
\hline
NuminaMath-CoT
& - & $\langle$4,1$\rangle\times$16 & $\langle$8,1$\rangle\times$8 \\
\hline
PubMedQA
& - & $\langle$4,1$\rangle\times$16 & $\langle$2,4$\rangle\times$8 \\
\hline
MetaMathQA
& - & $\langle$2,2$\rangle\times$16 & $\langle$4,2$\rangle\times$8 \\
\hline
Evol-Instruct
& $\langle$8,1$\rangle\times$2 & $\langle$8,1$\rangle\times$8 & $\langle$16,1$\rangle\times$4 \\
\hline
cnn\_dailymail
& - & $\langle$4,1$\rangle\times$16 & $\langle$8,1$\rangle\times$8 \\
\hline
XSum
& $\langle$8,1$\rangle\times$2 & $\langle$8,1$\rangle\times$8 & $\langle$16,1$\rangle\times$4 \\
\hline
MeetingBank
& $\langle$8,1$\rangle\times$2 & $\langle$8,1$\rangle\times$8 & $\langle$16,1$\rangle\times$4 \\ 
\hline
\end{tabular}
\end{table}

\begin{table}[H]
\centering
\captionof{table}{\small{Parallel configurations used by \system-Sequential for the end-to-end evaluation. (Those used by Task-Fused and \system are provided in Table~\ref{tb:expr_e2e_parallel_config}.) Each $\langle\alpha,\beta\rangle\times\gamma$ indicates that there are $\gamma$ FT replica(s) with a TP degree of $\alpha$ and a PP degree of $\beta$.}}
\label{tb:appendix_lobra_sequential_config}
\begin{tabular}{|c||c|c|c|}
\hline
\multirow{2}*{\textbf{Dataset Name}} & \multicolumn{3}{c|}{\textbf{\system-Sequential}} \\
\cline{2-4}
& \textbf{7B} & \textbf{32B} & \textbf{70B} \\ \hline 
\hline
MathInstruct 
& - & $\langle$1,2$\rangle\times$2, $\langle$2,1$\rangle\times$26, $\langle$1,4$\rangle\times$1, $\langle$2,2$\rangle\times$1 & $\langle$2,4$\rangle\times$5, $\langle$4,1$\rangle\times$2, $\langle$4,2$\rangle\times$2 \\
\hline
python\_code\_instructions
& $\langle$2,1$\rangle\times$2, $\langle$4,1$\rangle\times$1, $\langle$8,1$\rangle\times$1
& $\langle$1,2$\rangle\times$2, $\langle$2,1$\rangle\times$2, $\langle$4,1$\rangle\times$10, $\langle$8,1$\rangle\times$2
& $\langle$2,4$\rangle\times$1, $\langle$4,1$\rangle\times$2, $\langle$8,1$\rangle\times$2, $\langle$16,1$\rangle\times$2 \\
\hline
databricks-dolly-15k
& $\langle$1,1$\rangle\times$4, $\langle$1,4$\rangle\times$2, $\langle$4,1$\rangle\times$1
& $\langle$1,4$\rangle\times$2, $\langle$2,1$\rangle\times$6, $\langle$2,2$\rangle\times$9, $\langle$4,1$\rangle\times$2
& $\langle$2,4$\rangle\times$2, $\langle$4,1$\rangle\times$4, $\langle$4,2$\rangle\times$3, $\langle$8,1$\rangle\times$1 \\ 
\hline
BillSum
& -
& $\langle$2,2$\rangle\times$9, $\langle$4,1$\rangle\times$3, $\langle$8,1$\rangle\times$2
& $\langle$8,1$\rangle\times$2, $\langle$16,1$\rangle\times$3 \\
\hline
CommitPackFt
& $\langle$1,1$\rangle\times$10, $\langle$2,1$\rangle\times$1, $\langle$4,1$\rangle\times$1
& $\langle$2,2$\rangle\times$12, $\langle$4,1$\rangle\times$4
& $\langle$2,4$\rangle\times$1, $\langle$4,1$\rangle\times$6, $\langle$8,1$\rangle\times$4 \\ 
\hline
NuminaMath-CoT
& -
& $\langle$1,4$\rangle\times$4, $\langle$2,1$\rangle\times$20, $\langle$4,1$\rangle\times$2
& $\langle$2,4$\rangle\times$3, $\langle$4,1$\rangle\times$6, $\langle$4,2$\rangle\times$1, $\langle$8,1$\rangle\times$1 \\
\hline
PubMedQA
& -
& $\langle$1,2$\rangle\times$10, $\langle$2,1$\rangle\times$14, $\langle$1,4$\rangle\times$4
& $\langle$2,2$\rangle\times$10, $\langle$4,1$\rangle\times$4, $\langle$2,4$\rangle\times$1 \\
\hline
MetaMathQA
& -
& $\langle$1,4$\rangle\times$14, $\langle$2,1$\rangle\times$2, $\langle$2,2$\rangle\times$1
& $\langle$2,4$\rangle\times$6, $\langle$4,1$\rangle\times$2, $\langle$4,2$\rangle\times$1 \\
\hline
Evol-Instruct
& $\langle$1,1$\rangle\times$6, $\langle$2,1$\rangle\times$1, $\langle$8,1$\rangle\times$1
& $\langle$2,1$\rangle\times$16, $\langle$2,2$\rangle\times$1, $\langle$4,1$\rangle\times$3, $\langle$8,1$\rangle\times$2
& $\langle$2,4$\rangle\times$1, $\langle$4,1$\rangle\times$4, $\langle$8,1$\rangle\times$3, $\langle$16,1$\rangle\times$1 \\
\hline
cnn\_dailymail
& -
& $\langle$1,4$\rangle\times$1, $\langle$2,1$\rangle\times$14, $\langle$4,1$\rangle\times$8
& $\langle$2,4$\rangle\times$6, $\langle$4,2$\rangle\times$1, $\langle$8,1$\rangle\times$1 \\
\hline
XSum
& $\langle$1,1$\rangle\times$6, $\langle$2,1$\rangle\times$1, $\langle$8,1$\rangle\times$1
& $\langle$1,2$\rangle\times$8, $\langle$4,1$\rangle\times$8, $\langle$8,1$\rangle\times$2
& $\langle$2,4$\rangle\times$2, $\langle$4,1$\rangle\times$2, $\langle$4,2$\rangle\times$1, $\langle$16,1$\rangle\times$2 \\
\hline
MeetingBank
& $\langle$1,4$\rangle\times$2, $\langle$8,1$\rangle\times$1
& $\langle$1,2$\rangle\times$6, $\langle$1,4$\rangle\times$3, $\langle$8,1$\rangle\times$5
& $\langle$4,2$\rangle\times$1, $\langle$8,1$\rangle\times$1, $\langle$16,1$\rangle\times$3 \\ 
\hline
\end{tabular}
\end{table}

\begin{figure}[H]
\begin{minipage}[t]{0.47\linewidth}
\centering
\captionof{table}{\small{Parallel configurations used by Task-Fused and \system in Task scalability evaluation.}}
\label{tb:appendix_task_scalability_config}
\begin{tabular}{|c|c|c|}
\hline
\textbf{\# Tasks} & \textbf{Task-Fused} & \textbf{\system} \\ \hline \hline
4 & $\langle$16,1$\rangle\times$4 & $\langle$2,4$\rangle\times$3, $\langle$4,2$\rangle\times$1, $\langle$8,1$\rangle\times$2, $\langle$16,1$\rangle\times$1 \\ \hline
8 & $\langle$16,1$\rangle\times$4 & $\langle$2,4$\rangle\times$3, $\langle$4,2$\rangle\times$2, $\langle$8,1$\rangle\times$1, $\langle$16,1$\rangle\times$1 \\ \hline
12 & $\langle$16,1$\rangle\times$4 & $\langle$2,4$\rangle\times$3, $\langle$4,2$\rangle\times$2, $\langle$8,1$\rangle\times$1, $\langle$16,1$\rangle\times$1 \\ \hline
16 & $\langle$16,1$\rangle\times$4 & $\langle$2,4$\rangle\times$3, $\langle$4,2$\rangle\times$2, $\langle$8,1$\rangle\times$1, $\langle$16,1$\rangle\times$1 \\ \hline
\end{tabular}
\begin{minipage}[t]{0.05\linewidth}
$ $ 
\end{minipage}
\end{minipage}
\begin{minipage}[t]{0.47\linewidth}
\centering
\captionof{table}{\small{Parallel configurations used by Task-Fused and \system in GPU scalability evaluation.}}
\label{tb:appendix_gpu_scalability_config}
\begin{tabular}{|c|c|c|}
\hline
\textbf{\# GPUs} & \textbf{Task-Fused} & \textbf{\system} \\ \hline \hline
16 & $\langle$16,1$\rangle\times$1 & $\langle$16,1$\rangle\times$1 \\ \hline
32 & $\langle$16,1$\rangle\times$2 & $\langle$2,4$\rangle\times$1, $\langle$4,2$\rangle\times$1, $\langle$16,1$\rangle\times$1 \\ \hline
64 & $\langle$16,1$\rangle\times$4 & $\langle$2,4$\rangle\times$3, $\langle$4,2$\rangle\times$1, $\langle$8,1$\rangle\times$2, $\langle$16,1$\rangle\times$1 \\ \hline
\end{tabular}
\end{minipage}
\end{figure}

\clearpage

\section{System Comparison with Homogeneous Configurations}
\label{sec:appendix_nemo_compare}
As mentioned in \S\ref{sec:expr_setup}, \system achieves comparable efficiency against NeMo~\cite{nvidia_nemo}, a popular LLM training framework developed by NVIDIA, when training with homogeneous FT replicas. 
Here we present the results of extensive experiments to compare the performance of \system and NeMo thoroughly. 
To be specific, we consider fine-tuning the 7B model over 16 A100-40GB GPUs with a global batch size of 64 and varying the sequence lengths in different experiments. 
For each experiment, \system and NeMo utilize the same homogeneous parallel configuration and uniform data dispatching for fair comparison. 
In addition, to eliminate the impact of dynamic bucketing, all training data (sequences) are truncated or padded to the same maximum sequence length. 
The number of chunks (i.e. micro-batches) for gradient accumulation is tuned to achieve the best efficiency whilst avoiding out-of-memory errors. 
The results, summarized in Table \ref{tb:appendix_nemo_compare}, demonstrate that \system delivers comparable performance against NeMo when training with the same homogeneous parallel configurations.

\begin{table}[H]
\centering
\caption{\small{System comparison of \system and NeMo with different homogeneous parallel configurations.}}
\label{tb:appendix_nemo_compare}
\begin{tabular}{|c|c|c|c|c|c|c|c|}
\hline
\textbf{\# GPUs} & \textbf{Config} & \textbf{Max Sequence Length} & \textbf{\# Chunks (Micro Batches)} & \textbf{\system Time (s)} & \textbf{NeMo Time (s)} \\ \hline \hline
16 & $\langle$TP=1,PP=1$\rangle\times$16 & 2048 & 4 & 1.778 & 1.533 \\ \hline
16 & $\langle$TP=1,PP=2$\rangle\times$8 & 2048 & 8 & 1.978 & 1.785 \\ \hline
16 & $\langle$TP=1,PP=4$\rangle\times$4 & 2048 & 16 & 2.131 & 1.939 \\ \hline
16 & $\langle$TP=1,PP=4$\rangle\times$4 & 4096 & 16 & 4.141 & 3.872 \\ \hline
16 & $\langle$TP=1,PP=8$\rangle\times$2 & 2048 & 32 & 2.308 & 2.134 \\ \hline
16 & $\langle$TP=1,PP=8$\rangle\times$2 & 4096 & 32 & 4.492 & 4.247 \\ \hline
16 & $\langle$TP=2,PP=1$\rangle\times$8 & 2048 & 8 & 2.414 & 2.127 \\ \hline
16 & $\langle$TP=2,PP=1$\rangle\times$8 & 4096 & 8 & 4.297 & 3.922 \\ \hline
16 & $\langle$TP=2,PP=2$\rangle\times$4 & 2048 & 16 & 2.611 & 2.432 \\ \hline
16 & $\langle$TP=2,PP=2$\rangle\times$4 & 4096 & 16 & 4.612 & 4.294 \\ \hline
16 & $\langle$TP=2,PP=4$\rangle\times$2 & 2048 & 32 & 2.718 & 2.616 \\ \hline
16 & $\langle$TP=2,PP=4$\rangle\times$2 & 4096 & 32 & 4.915 & 4.548 \\ \hline
16 & $\langle$TP=2,PP=8$\rangle\times$1 & 2048 & 64 & 3.040 & 2.967 \\ \hline
16 & $\langle$TP=2,PP=8$\rangle\times$1 & 4096 & 64 & 5.391 & 5.221 \\ \hline
16 & $\langle$TP=2,PP=8$\rangle\times$1 & 8192 & 64 & 10.611 & 9.956 \\ \hline
16 & $\langle$TP=4,PP=1$\rangle\times$4 & 2048 & 16 & 3.395 & 4.040 \\ \hline
16 & $\langle$TP=4,PP=1$\rangle\times$4 & 4096 & 16 & 5.608 & 5.198 \\ \hline
16 & $\langle$TP=4,PP=1$\rangle\times$4 & 8192 & 16 & 10.530 & 9.956 \\ \hline
16 & $\langle$TP=4,PP=2$\rangle\times$2 & 2048 & 32 & 3.626 & 4.447 \\ \hline
16 & $\langle$TP=4,PP=2$\rangle\times$2 & 4096 & 32 & 5.911 & 5.494 \\ \hline
16 & $\langle$TP=4,PP=2$\rangle\times$2 & 8192 & 32 & 11.143 & 10.634 \\ \hline
16 & $\langle$TP=4,PP=4$\rangle\times$1 & 2048 & 64 & 3.793 & 4.637 \\ \hline
16 & $\langle$TP=4,PP=4$\rangle\times$1 & 4096 & 64 & 6.255 & 5.939 \\ \hline
16 & $\langle$TP=4,PP=4$\rangle\times$1 & 8192 & 64 & 11.770 & 11.139 \\ \hline
16 & $\langle$TP=8,PP=1$\rangle\times$2 & 2048 & 32 & 5.691 & 8.494 \\ \hline
16 & $\langle$TP=8,PP=1$\rangle\times$2 & 4096 & 32 & 8.649 & 8.589 \\ \hline
16 & $\langle$TP=8,PP=1$\rangle\times$2 & 8192 & 32 & 14.769 & 13.770 \\ \hline
16 & $\langle$TP=8,PP=1$\rangle\times$2 & 16384 & 32 & 29.271 & 28.054 \\ \hline
16 & $\langle$TP=8,PP=2$\rangle\times$1 & 2048 & 64 & 5.887 & 8.693 \\ \hline
16 & $\langle$TP=8,PP=2$\rangle\times$1 & 4096 & 64 & 9.032 & 9.028 \\ \hline
16 & $\langle$TP=8,PP=2$\rangle\times$1 & 8192 & 64 & 15.464 & 14.669 \\ \hline
16 & $\langle$TP=8,PP=2$\rangle\times$1 & 16384 & 64 & 30.468 & 29.299 \\ \hline
\end{tabular}
\end{table}

\clearpage

\section{Explanation of Time Cost Modeling}
\label{sec:appendix_time_cost_model}

The results on the right side of Figure \ref{fig:expr_effectiveness} demonstrate that our cost model is accurate. 
However, due to the space constraint, \S\ref{sec:pre_parallel_config} does not present the details of our cost model. 
Here we explain its design in detail.

Following previous works~\cite{galvatron,alpa}, our cost model is built on top of offline profiling and curve fitting. For an input batch of size $(b, s)$, where $b$ and $s$ represent the batch size and sequence length respectively, the training time $t(b, s)$ is composed of forward and backward passes. The running time for each pass is proportional to $b$, and from the perspective of modules, the running time of the attention mechanism is proportional to the square of $s$, whilst for other modules, it is proportional to $s$. Therefore, we construct the cost model as a function that is quadratic with respect to $s$ and proportional to $b$. This model is fitted using offline profiling data for various $(b, s)$ pairs, enabling us to accurately estimate the forward and backward times for an input batch. Furthermore, since modern large models (e.g., LLMs) typically consist of identical layers, we simplify and expedite the offline process by profiling only a single layer.

In practice, due to the memory limit and/or the use of pipeline parallel, training usually consists of multiple chunks (i.e., micro-batches), so it is necessary to construct a time cost function for the overall training process. We introduce our time cost function design in two stages: with and without pipeline parallel.

For training without pipeline parallel, the time cost is solely composed of forward and backward computation. Given $R$ sequence buckets and $d_{j}$ sequences of length $s_{j}$ in the $j$-th bucket, for parallel configuration $\mathcal{S}$ with a maximum supportable sequence length $M$, the time cost function can be formulated as:
\begin{equation}
\label{eq:appendix_cost_model_w/o_pp}
    \begin{aligned}
        T(\{d_j\}_{j=1}^{R};\mathcal{S}) &= \sum_{j=1}^{R}{\left(m_j \cdot t(b_j, s_j) + t(r_j, s_j)\right)} \\
        \text{where}\ \ \ d_j &= m_j \cdot b_j + r_j \\
        b_j &= \lfloor \frac{M}{s_j} \rfloor \\
    \end{aligned}
\end{equation}

When pipeline parallel is employed, the training batch is evenly divided into multiple micro-batches, and the forward and backward passes across micro-batches are executed on different stages in a pipeline manner. This process is composed of three stages: warm-up, steady, and cool-down. The time cost includes additional overhead such as data transfer and synchronization between pipeline stages. Specifically, in 1F1B pipeline parallel with $p$ stages, if a mini-batch of size $(b, s)$ is evenly split into $m$ micro-batches, the time cost can be formulated as:
\begin{equation}
\label{eq:appendix_cost_model_pp_fixed_len}
    T(b, s, m) = \underbrace{m \cdot t(\frac{b}{m}, s)}_{\text{Compute Time}} + \underbrace{(p - 1) \cdot t(\frac{b}{m}, s)}_{\text{Bubble Time}}
\end{equation}
where the former represents computation time and the latter represents pipeline bubble time.

However, dynamic bucketing introduces challenges due to variable-length input. As presented in Figure \ref{fig:critical_path}, the critical path of a variable-length pipeline does not have a fixed paradigm as fixed-length pipeline parallel where the length is consistent across micro-batches, making precise time cost estimation more complicated. Additionally, variable-length pipelines introduce additional bubbles caused by the imbalanced time costs of micro-batches, resulting in increased idle time.\footnote{In fact, reducing pipeline bubbles for the training of variable-length data is important and meaningful work. However, it is orthogonal to the goal of our work. And any optimization regarding the reduction in pipeline bubbles can be integrated with our work.}

\begin{figure}[H]
\centering
\includegraphics{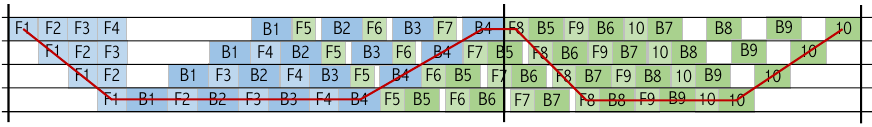}
\caption{\small{Illustration of the 1F1B pipeline parallel execution with variable-length inputs. Cells in light background color and dark background color represent forward and backward processes, respectively. The blue and green indicate two kinds of micro-batches that are different in length. The critical path of the execution is highlighted in red.}}
\label{fig:critical_path}
\end{figure}

To reduce fragmented bubble time caused by variable-length inputs and to simplify time cost estimation, we propose sorting micro-batches in descending order of time cost and partitioning the critical path based on length for phased estimation. For the first and the most time-consuming micro-batches, the critical path includes the warm-up, steady, and cool-down stages of these micro-batches, which we define as the first phase. For the remaining micro-batches, the critical path is partitioned according to their length.

As shown in Figure \ref{fig:critical_path}, the time cost of the first phase is expressed as $m_1 \cdot t(b_1, s_1) + (p-1) \cdot t(b_1, s_1)$, accounting for both the compute time of micro-batches and the bubble time estimation of the entire pipeline.
Since micro-batches with different lengths have interleaving characteristics in the pipeline, the time cost of each remaining phase does not exceed the computation time of micro-batches with the same length, formulated as $m_i \cdot t(b_i, s_i)$.

Putting the above together, to effectively estimate the time cost for the variable-length case, we formulate the time cost function as follows:
\begin{equation}
\label{eq:appendix_cost_model_pp_varlen}
    \begin{aligned}
        T(\{d_j\}_{j=1}^{R};\mathcal{S}) &= \underbrace{\sum_{i=0}^{R-1}{\left(m_i \cdot t(b_i, s_i) + t(r_i, s_i)\right)}}_{\text{Compute Time}} \\
        &+ \underbrace{(p - 1) \times \max_{0 \le j \le R-1}{\{t(b_j, s_j), t(r_j, s_j)\}}}_{\text{Bubble Time}} \\
        \text{where}\ \ \ n_i &= m_i \cdot b_i + r_i \\
        b_i &= \lfloor \frac{M}{s_i} \rfloor
    \end{aligned}
\end{equation}
It is worth noting that our time cost function is compatible with the previously mentioned time cost function without pipeline parallel and with fixed-length pipeline parallel.

Last but not least, we wish to highlight that our work is applicable as long as the time cost function $T(\{d_{\cdot, j}\}_{j=1}^{r};\mathcal{S})\ $is linear w.r.t. $d_{\cdot, j}$. Either solving the ILP or MINLP problems mentioned in \S\ref{sec:method_deployment} and \S\ref{sec:method_dispatching}, $\{d_{\cdot, j}\}_{j=1}^{r_j}$ serve as the decision variables. Meanwhile, the objective is to minimize the maximum time of all FT replicas, presented as objective variable $\hat{t} \ge T\left( \left\{ \left\lceil \frac{d_{i,j}}{p_i} \right\rceil \right\}_{j=1}^{r_i}; \mathcal{S}_i \right)$ that holds for all configurations $\{\mathcal{S}_i\}_{i=1}^S$. The linearity between the time cost function and $\{d_{\cdot, j}\}$ ensures that the constraint of the objective variable $\hat{t}$ remains linear, thereby preserving the ILP or MINLP properties of the problem formulation.

\fi 

\end{document}